\documentclass[twocolumn]{aastex631}

\usepackage[normalem]{ulem}

\submitjournal{ApJ}

\shorttitle{Chemical compositions of the vicinity of protostars in Ophiuchus}
\shortauthors{Taniguchi et al.}
\begin{document}

\title{Chemical compositions in the vicinity of protostars in Ophiuchus}

\correspondingauthor{Kotomi Taniguchi; Liton Majumdar}
\email{kotomi.taniguchi@nao.ac.jp; liton@niser.ac.in}

\author[0000-0003-4402-6475]{Kotomi Taniguchi}
\affiliation{National Astronomical Observatory of Japan, National Institutes of Natural Sciences, 2-21-1 Osawa, Mitaka, Tokyo 181-8588, Japan}

\author[0000-0001-7031-8039]{Liton Majumdar}
\affiliation{School of Earth and Planetary Sciences, National Institute of Science Education and Research, HBNI, Jatni 752050, Odisha, India}

\author[0000-0002-9912-5705]{Adele Plunkett}
\affiliation{National Radio Astronomy Observatory, 520 Edgemont Rd., Charlottesville, VA 22903, USA}

\author[0000-0003-0845-128X]{Shigehisa Takakuwa}
\affiliation{Department of Physics and Astronomy, Graduate School of Science and Engineering, Kagoshima University, 1-21-35 Korimoto, Kagoshima, Kagoshima 890-0065, Japan}

\author[0000-0002-0500-4700]{Dariusz C. Lis}
\affiliation{Jet Propulsion Laboratory, California Institute of Technology, 48010 Oak Grove Drive, Pasadena, CA 91109, USA}

\author[0000-0002-6622-8396]{Paul F. Goldsmith}
\affiliation{Jet Propulsion Laboratory, California Institute of Technology, 48010 Oak Grove Drive, Pasadena, CA 91109, USA}

\author[0000-0001-5431-2294]{Fumitaka Nakamura}
\affiliation{National Astronomical Observatory of Japan, National Institutes of Natural Sciences, 2-21-1 Osawa, Mitaka, Tokyo 181-8588, Japan}
\affiliation{Department of Astronomical Science, School of Physical Science, SOKENDAI (The Graduate University for Advanced Studies), Osawa, Mitaka, Tokyo 181-8588, Japan}
\affiliation{The University of Tokyo, Hongo, Bunkyo, Tokyo 113-0033, Japan}

\author[0000-0003-0769-8627]{Masao Saito}
\affiliation{National Astronomical Observatory of Japan, National Institutes of Natural Sciences, 2-21-1 Osawa, Mitaka, Tokyo 181-8588, Japan}
\affiliation{Department of Astronomical Science, School of Physical Science, SOKENDAI (The Graduate University for Advanced Studies), Osawa, Mitaka, Tokyo 181-8588, Japan}

\author[0000-0002-4649-2536]{Eric Herbst}
\affiliation{Department of Astronomy, University of Virginia, Charlottesville, VA 22904, USA}
\affiliation{Department of Chemistry, University of Virginia, Charlottesville, VA 22904, USA}



\begin{abstract}
We have analyzed Atacama Large Millimeter/submillimeter Array (ALMA) Cycle 4 Band 6 data toward two young stellar objects (YSOs), Oph-emb5 and Oph-emb9, in the Ophiuchus star-forming region.
The YSO Oph-emb5 is located in a relatively quiescent region, whereas Oph-emb9 is irradiated by a nearby bright Herbig Be star.
Molecular lines from $cyclic$-C$_{3}$H$_{2}$ ($c$-C$_{3}$H$_{2}$), H$_{2}$CO, CH$_{3}$OH, $^{13}$CO, C$^{18}$O, and DCO$^{+}$ have been detected from both sources, while DCN is detected only in Oph-emb9.
Around Oph-emb5, $c$-C$_{3}$H$_{2}$ is enhanced at the west side, relative to the IR source, whereas H$_{2}$CO and CH$_{3}$OH are abundant at the east side.
In the field of Oph-emb9, moment 0 maps of the $c$-C$_{3}$H$_{2}$ lines show a peak at the eastern edge of the field of view, which is irradiated by the Herbig Be star.
Moment 0 maps of CH$_{3}$OH and H$_{2}$CO show peaks farther from the bright star.
We derive the $N$($c$-C$_{3}$H$_{2}$)/$N$(CH$_{3}$OH) column density ratios at the peak positions of $c$-C$_{3}$H$_{2}$ and CH$_{3}$OH near each YSO, which are identified based on their moment 0 maps.
The $N$($c$-C$_{3}$H$_{2}$)/$N$(CH$_{3}$OH) ratio at the $c$-C$_{3}$H$_{2}$ peak is significantly higher than at the CH$_{3}$OH peak by a factor of $\sim 19$ in Oph-emb9, while the difference in this column density ratio between these two positions is a factor of $\sim2.6 $ in Oph-emb5.
These differences are attributed to the efficiency of the photon-dominated region (PDR) chemistry in Oph-emb9.
The higher DCO$^{+}$ column density and the detection of DCN in Oph-emb9 are also discussed in the context of UV irradiation flux.
\end{abstract}



\section{Introduction} \label{sec:intro}

Chemical composition is a powerful tool with which to investigate the physical conditions and their evolution \citep[]{2012A&ARv..20...56C} from core scale ($< 0.1$ pc) to clump scale in molecular clouds \citep[$\sim1$ pc;][]{2020MNRAS.493.2395T}.
These studies around low-mass protostars are essential for understanding the formation processes of our solar system and complex organic molecules detected in solar system bodies \citep{2017ApJ...850..176C}. %Two different types of molecules are formed around protostars.
%One consists of complex organic molecules (COMs), which consist of more than six atoms \citep{2009ARA&A..47..427H}.
%A fundamental COM is methanol (CH$_{3}$OH), which is thought to be mainly formed by successive hydrogenation reactions of CO molecules on dust grain surfaces \citep{2002ApJ...571L.173W}.
%Radicals which are formed by irradiation of CH$_{3}$OH form even more complex COMs \citep{2021PhR...893....1O}.
%Hot ($> 100$ K) and dense ($\geq 10^{6}$ cm$^{-3}$) gaseous regions rich in COMs are called hot cores and hot corinos in high-mass and low-mass YSOs, respectively.
%The other type of molecule is unsaturated carbon-chain molecules.
%Carbon-chain molecules around YSOs can be formed from CH$_{4}$, which sublimates from dust grains around 25--35 K \citep{2008ApJ...681.1385H,2019ApJ...881...57T}.
%Such a carbon-chain formation mechanism was first discovered in the L1527 low-mass YSO \citep{2008ApJ...672..371S}. Carbon-chain formation originating from CH$_{4}$ was named warm carbon-chain chemistry \citep[WCCC;][]{2008ApJ...672..371S}.
%Recent observations toward high-mass YSOs also revealed that carbon-chain chemistry may also occur in high temperature regions \citep{2018ApJ...866..150T,2021ApJ...908..100T}.
Chemical differentiation around low-mass young stellar objects (YSOs) has been proposed since the late 2000s.
One such chemical  process is called hot-corino chemistry, which leads to abundant complex organic molecules (COMs) consisting of more than six atoms \citep{2009ARA&A..47..427H}.
The other important one is called warm carbon-chain chemistry \citep[WCCC;][]{2008ApJ...672..371S}.
Recent observations have shown chemical differentiation not only around low-mass YSOs, but around high-mass YSOs as well \citep{2018ApJ...866..150T, 2019ApJ...872..154T, 2021ApJ...908..100T}.
Although the origin of the chemical differentiation around YSOs is still controversial, three possible factors have been proposed: the different timescale of the prestellar collapse \citep{2008ApJ...672..371S}, the different ultraviolet (UV) radiation field \citep{2016A&A...586A.110S}, and the different timescale of the warm-up stage \citep{2019ApJ...881...57T}. In order to reveal the effects of the above factors, we need to investigate molecular spatial distributions on large scales (e.g., the clump scale).

Our Sun is considered to have been born as a member of a cluster \citep{2010ARA&A..48...47A, 2019A&A...631A..25J}.
In such a region, nearby sources may play important roles in chemical processes. 
One of the processes affecting chemical composition is UV radiation, as was suggested by \citet{2016A&A...586A.110S} based on observations toward starless cores.
They found that CH$_{3}$OH is abundant in a region well shielded against the interstellar radiation field, while $cyclic$-C$_{3}$H$_{2}$ (hereafter $c$-C$_{3}$H$_{2}$) is enhanced in the irradiated environment.
Such a chemical differentiation implies a different chemical composition in ice mantles, as shown by the CH$_{4}$/CH$_{3}$OH ice-mantle abundance ratio \citep{2016A&A...586A.110S, 2020A&A...643A..60S}, because the gas-phase $c$-C$_{3}$H$_{2}$ can be considered to form from CH$_{4}$ by the WCCC mechanism \citep{2008ApJ...681.1385H}.
The UV radiation destroys CO molecules forming carbon atoms (C), which lead to the CH$_{4}$-rich ice, and finally the carbon-chain-rich gas.

Chemical differentiation around low-mass YSOs has been studied using single-dish telescopes.
\citet{2016ApJ...833L..14L} carried out survey observations of C$_{4}$H and CH$_{3}$OH toward 16 low-mass YSOs in the Ophiuchus and Corona Australis molecular clouds using the Kitt Peak 12-m radio telescope and APEX.
They proposed a spatial separation between these two types of molecules.
Another study by \citet{2017ApJ...835....3L} presented APEX observations of H$_{2}$CO and $c$-C$_{3}$H$_{2}$ toward protostars in the Ophiuchus star-forming region.
They suggested that the $c$-C$_{3}$H$_{2}$ emission traces the more shielded parts of the envelope, whereas the H$_{2}$CO emission mainly traces the outer irradiated envelopes.
Their suggestion seems to be opposite to the finding in starless cores \citep{2016A&A...586A.110S, 2020A&A...643A..60S}.
Our target YSOs, identified as Oph-emb5 and Oph-emb9 in this paper, were also observed by \citet{2016ApJ...833L..14L, 2017ApJ...835....3L}, and carbon-chain species (C$_{4}$H and $c$-C$_{3}$H$_{2}$) and COMs (CH$_{3}$OH and H$_{2}$CO) have been detected from both YSOs.
These studies \citep{2016ApJ...833L..14L,2017ApJ...835....3L}, however, were single-dish single-pointing observations, and spatial variation among these molecular lines were not resolved.
High-angular resolution and high sensitivity observations are needed to study the effects of UV radiation on chemical composition around YSOs in order to distinguish between nearby sources and target YSOs.

In this paper, we report ALMA Band 6 data toward two YSOs in the Ophiuchus region. 
The Ophiuchus region is a nearby star-forming region \citep[$\sim140$ pc][]{2018ApJ...869L..33O} with Class II YSOs relatively more abundant than YSOs of other classes, similar to the Lupus I region and the Chamaeleon region \citep{2015ApJS..220...11D}.
A bright Herbig Be star (S1) irradiates the $\rho$ Oph A cloud \citep{2017ApJ...835....3L}.
The YSO Oph-emb9, one of our target sources, is located in this cloud and is irradiated by the Herbig Be star from the east.
The YSO Oph-emb5 is located in the $\rho$ Oph B cloud with no nearby irradiation sources. 
The distance between the $\rho$ Oph A cloud and the $\rho$ Oph B cloud is around 0.55 pc, and we can assume that the other initial conditions are comparable.
Thus, these are good target regions to study effects of nearby bright sources on the chemical differentiation around YSOs.

The structure of the present paper is as follows. In Section \ref{sec:data}, we explain the data sets and reduction procedure.
The resultant continuum maps are presented in Section \ref{sec:resmom0}, moment 0 maps of the detected molecular lines are shown in Section \ref{sec:3.3}, spectra and spectral analyses are presented in Section \ref{sec:specana}, and finally moment 2 maps of $^{13}$CO and C$^{18}$O are shown in Section \ref{sec:resmom2}. 
Our main conclusions are summarized in Section \ref{sec:con}.

\section{Observations and data reduction} \label{sec:data}

We have analyzed ALMA Band 6 archival data toward two YSOs in the Ophiuchus region taken as part of a Cycle 4 project\footnote{project ID; 2016.1.00319.S, PI: Johan Lindberg}.
Table \ref{tab:source} summarizes the coordinates and properties of our two target YSOs.
Based on the infrared spectral indices, Oph-emb5 and Oph-emb9 are classified as Flat SED\footnote{A definition and characteristics are summarized in \citet{2009arXiv0901.1691E}.} and Class I, respectively.

%Table 1
\begin{deluxetable*}{cccccccc}
\tablenum{1}
\tablecaption{Summary of Target YSOs\label{tab:source}}
\tablewidth{0pt}
\tablehead{
\colhead{Source} & \colhead{R.A. (J2000)\tablenotemark{a}} & \colhead{Decl. (J2000)\tablenotemark{a}} & \colhead{$L_{\rm {bol}}$ (L$_{\odot}$)\tablenotemark{b}} & \colhead{$T_{\rm {bol}}$ (K)\tablenotemark{c}} &\colhead{$M_{\rm {env}}$ (M$_{\odot}$)\tablenotemark{d}} &  \colhead{$\alpha_{\rm {IR}}$\tablenotemark{e}} & \colhead{Class\tablenotemark{f}} %\colhead{Other identifiers}
}
%\decimalcolnumbers
\startdata
Oph-emb5 & $16^{\rm {h}}27^{\rm {m}}$21\fs96 & -24\degr27\arcmin27\farcs7 & 0.1 (0.1) & 87 (28) & 0.15 (0.01) & -0.05 (0.05) & Flat  \\%& Oph-emb5, J162721.8-242727 \\
Oph-emb9 & $16^{\rm {h}}26^{\rm {m}}$25\fs44 & -24\degr23\arcmin01\farcs3 & 0.12 (0.37) & 135 (49) & 0.65 (0.05) & 0.87 (0.05) & I \\ %Oph-emb9, J162625.4-242301 \\
\enddata
\tablecomments{Numbers in parentheses indicate uncertainties corrected by the distances.}
\tablenotetext{a}{Coordinates of infrared sources determined by the Spitzer observations \citep{2009ApJ...692..973E}.}
\tablenotetext{b}{Bolometric luminosities at a distance of 125 pc taken from \citet{2009ApJ...692..973E} and scaled to the newly measured distances \citep[140.2 pc and 138.6 pc for Oph-emb5 and Oph-emb9, respectively;][]{2018ApJ...869L..33O}.}
\tablenotetext{c}{Bolometric temperature taken from \citet{2009ApJ...692..973E}.}
%Original values at 125 pc are 0.08 (0.08) and 0.1 (0.3) Lbol for Oph-emb5 and Oph-emb9, respectively.
\tablenotetext{d}{Envelope masses at a distance of 125 pc taken from \citet{2009ApJ...692..973E} and scaled to the newly measured distances \citep[140.2 pc and 138.6 pc for Oph-emb5 and Oph-emb9, respectively;][]{2018ApJ...869L..33O}.}
%Original values at 125 pc are 0.12 (0.01) and 0.53 (0.05) Msun for Oph-emb5 and Oph-emb9, respectively.
\tablenotetext{e}{IR spectral indexes ($\alpha_{\rm {IR}}$) taken from \citet{2009ApJ...692..973E}.}
\tablenotetext{f}{Classification taken from \citet{2015MNRAS.447.1996W}.}
\end{deluxetable*}

\begin{figure*}[!th]
\figurenum{1}
 \begin{center}
  \includegraphics[bb = 0 25 499 450, scale=0.55]{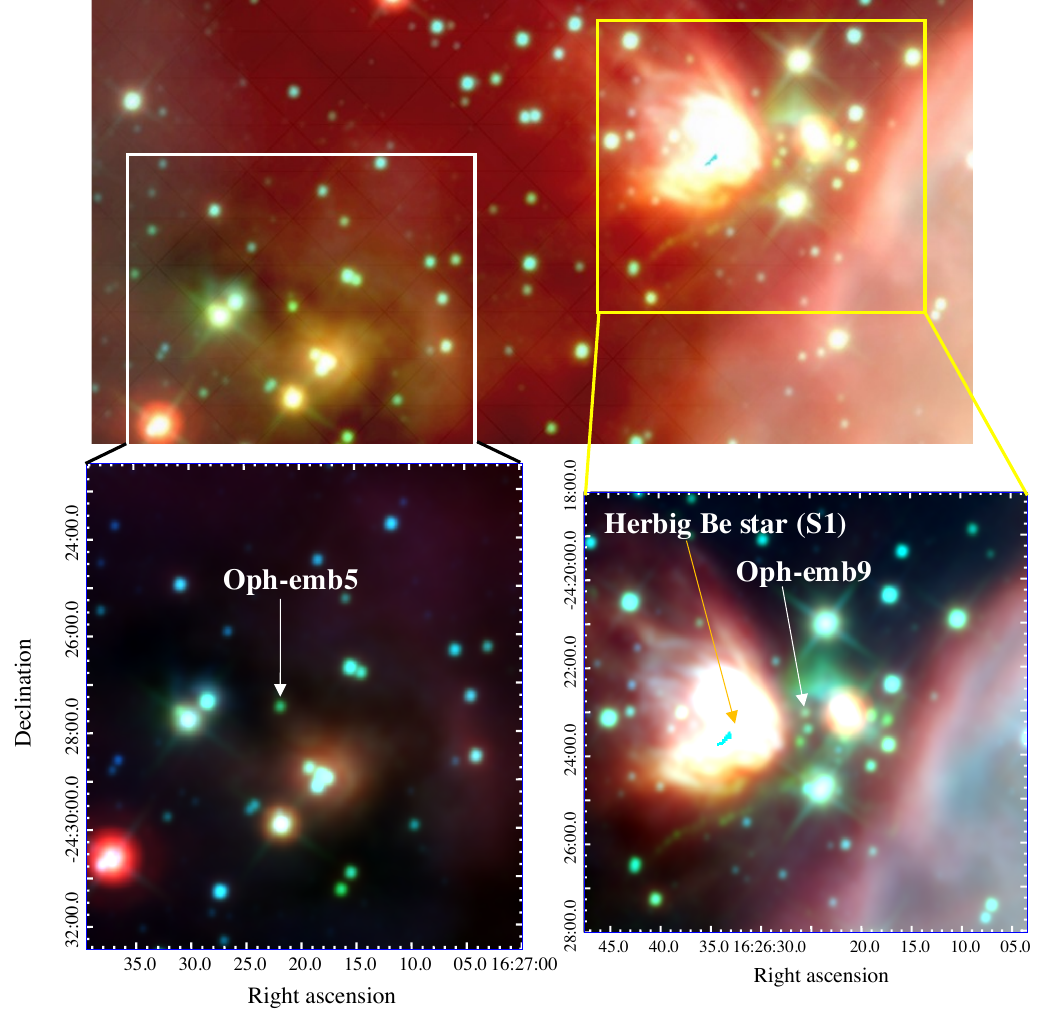}
 \end{center}
\caption{Wide-field Infrared Survey Explorer (WISE) images toward the two target YSOs; blue; W1 (3.4 $\mu$m), green; W2 (4.6 $\mu$m), red; W4 (22 $\mu$m), respectively. The color scales are different and have been adjusted for each panel. \label{fig:wise}}
\end{figure*}

Figure \ref{fig:wise} shows the Wide-field Infrared Survey Explorer \citep[WISE;][]{2010AJ....140.1868W} images toward the two target YSOs.
Oph-emb9 is irradiated by a luminous Herbig Be star S1 (also known as GSS35), located 2\arcmin ($\sim 16800$ au) east of the $\rho$ Oph A cloud \citep{2017ApJ...835....3L}.
This Herbig Be star has spectral class of B4 \citep{1992A&AS...92..481B} and luminosity of $\sim 1000-1600$ $L_{\odot}$ \citep{2001A&A...372..173B, 2005AJ....130.1733W}.
On the other hand, Oph-emb5 is located in a relatively isolated region and is not affected by any other sources. 

The data sets contain the 12-m array, 7-m array, and Total Power (TP) data.
The 12-m array and 7-m array data were obtained in 2017 March.
The TP observations were carried out in 2016 October toward Oph-emb9, and 2016 November and December, 2017 March and April toward Oph-emb5.
Coordinates of the phase reference centers are ($\alpha_{J2000}$, $\delta_{J2000}$) = ($16^{\rm {h}}27^{\rm {m}}$21\fs83, -24\degr27\arcmin27\farcs6) and ($16^{\rm {h}}26^{\rm {m}}$25\fs49, -24\degr23\arcmin01\farcs6) toward Oph-emb5 and Oph-emb9, respectively.

Table \ref{tab:Spw} summarizes the details of each spectral window.
The correlator setup with a frequency resolution of 61 kHz was used for molecular line observations.
This frequency resolution corresponds to a velocity resolution of $\sim 0.084$ km\,s$^{-1}$.
The velocity resolutions for each resultant cube are also summarized in Table \ref{tab:Spw}.
We employed a velocity resolution of 0.2 km\,s$^{-1}$, when lines can be resolved sufficiently by this velocity resolution.
The fields of views (FoV) of the 12-m array and 7-m array are $\sim27$\arcsec and $\sim46$\arcsec, respectively.

%Table2
\begin{deluxetable}{cccc}
\tablenum{2}
\tablecaption{Summary of spectral windows covered by the correlator setup\label{tab:Spw}}
\tablewidth{0pt}
\tablehead{
\colhead{Frequency Range} & \colhead{Molecule} & \colhead{Transition} & \colhead{$\Delta v$\tablenotemark{a}} \\
\colhead{(GHz)} & \colhead{} & \colhead{} & \colhead{(km\,s$^{-1}$)}
}
%\decimalcolnumbers
\startdata
217.92--217.96 & $c$-C$_{3}$H$_{2}$ & $5_{1, 4}- 4_{2, 3}$ & 0.084 \\
217.80--217.84 & $c$-C$_{3}$H$_{2}$ & $6_{0, 6}- 5_{1, 5}$ & 0.084 \\
218.20--218.24 & H$_{2}$CO & $3_{0, 3}-2_{0, 2}$ & 0.2 \\
218.45--218.49 & H$_{2}$CO & $3_{2, 2}- 2_{2, 1}$ & 0.084/0.2\tablenotemark{b} \\
218.74--218.78 & H$_{2}$CO & $3_{2, 1}- 2_{2, 0}$  & 0.084/0.2\tablenotemark{b} \\
218.42--218.46 & CH$_{3}$OH & $4_{-2, 3}- 3_{-1, 2}$ $E$ & 0.084/0.2\tablenotemark{b} \\
219.54--219.58 & C$^{18}$O & $2-1$ & 0.2 \\
220.38--220.41 & $^{13}$CO & $2-1$ & 0.2 \\
216.09--216.13 & DCO$^{+}$ & $3-2$ & 0.084 \\
217.22--217.26 & DCN & $3-2$ & 0.084 \\
216.0--218.0 & \multicolumn{2}{c}{Continuum} & ... \\ 
\enddata
\tablenotetext{a}{Velocity resolution of the resultant cubes.}
\tablenotetext{b}{The velocity resolution of 0.084 km\,s$^{-1}$ and 0.2 km\,s$^{-1}$ were applied for Oph-emb5 and Oph-emb9 data, respectively.}
\end{deluxetable}

We carried out data reduction and imaging using the Common Astronomy Software Application \citep[CASA][]{2007ASPC..376..127M} on the pipeline-calibrated visibilities.
We ran the calibration scripts using CASA version 4.7.0 for all of the data except for the TP data toward Oph-emb5, which was run with version 4.7.2.

The interferometric data cubes were created using the CASA ``tclean" task after concatenating, combining the 12-m array and 7-m array data by task ``concat''.
Briggs weighting with a robust parameter of 0.5 was applied.
The TP images were made using the ``sdimaging" task.
We combined the interferometer data with the TP data by using the feather task in CASA.
We applied an sdfactor of 1.2.
We conducted the primary beam correction for the combined images.
The resulting angular resolutions are approximately 2\farcs2 $\times$ 1\farcs5 and 1\farcs8 $\times$ 1\farcs3 in Oph-emb5 and Oph-emb9, respectively.
These angular resolutions correspond to 308 au $\times$ 210 au in Oph-emb5 and 249 au $\times$ 180 au in Oph-emb9 at the source distances \citep[140.2 pc and 138.6 pc for Oph-emb5 and Oph-emb9, respectively;][]{2018ApJ...869L..33O}.

Continuum images ($\lambda=1.38$ mm) with the 12-m array data were made by tclean task with the specmode of ``mfs'' in CASA.
The center frequency and band width for these continuum data are 217.0 GHz and 2 GHz, respectively.
The resulting angular resolutions are 2\farcs2 $\times$ 1\farcs5 and 1\farcs8 $\times$ 1\farcs3 for Oph-emb5 and Oph-emb9, respectively.
The noise levels of the continuum images are $6.0 \times 10^{-2}$ mJy\,beam$^{-1}$ and 1.0 mJy\,beam$^{-1}$ for Oph-emb5 and Oph-emb9, respectively.

\section{Results and Analyses} \label{sec:res}

\subsection{Continuum images} \label{sec:resmom0}

\begin{figure*}[!th]
\figurenum{2}
 \begin{center}
  \includegraphics[bb = 0 25 538 290, scale=0.65]{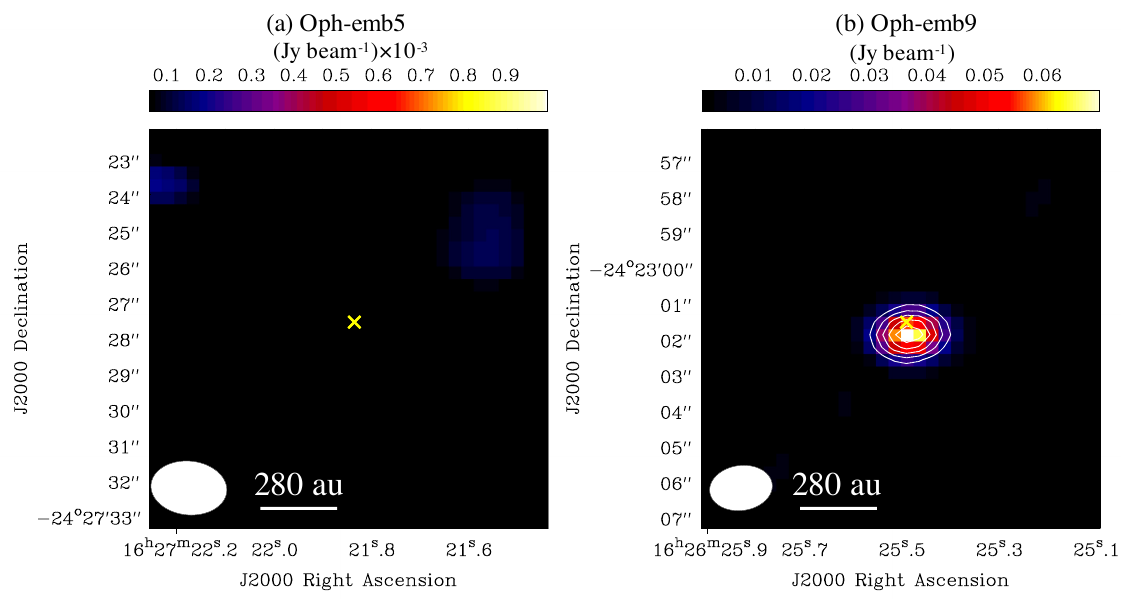}
 \end{center}
\caption{Continuum images ($\lambda=1.38$ mm) toward (a) Oph-emb5 and (b) Oph-emb9, respectively. The contour levels are 20, 40, 50, 60 times of the rms noise level, which are $6.0 \times 10^{-2}$ mJy\,beam$^{-1}$ and 1.0 mJy\,beam$^{-1}$ for panels (a) and (b), respectively. The yellow crosses indicate positions of the infrared sources \citep{2009ApJ...692..973E}. The filled white ellipses indicate the beam sizes of 2\farcs2 $\times$ 1\farcs5 and 1\farcs8 $\times$ 1\farcs3 for panels (a) and (b), respectively.\label{fig:cont}}
\end{figure*}

The continuum images toward the two YSOs are shown in Figure \ref{fig:cont}.
The yellow crosses indicate the positions of infrared (IR) sources identified by the Spitzer Core to Disk (c2d) Legacy program \citep{2009ApJ...692..973E}.
Continuum emission is detected toward the YSO Oph-emb9.
On the other hand, no continuum emission was detected toward the YSO Oph-emb5.
\citet{2019ApJ...871...86K} also did not detect the continuum emission toward this source (J162721 in their paper) at the 1.3 mm wavelength using ALMA.
The non-detection of dust continuum emission has also been reported toward other Class I YSOs \citep[e.g.,][]{2021ApJ...910..141T}.

\subsection{Moment 0 maps of the observed molecular lines} \label{sec:3.3}

\begin{figure*}[!th]
\figurenum{3}
 \begin{center}
  \includegraphics[bb = 0 30 645 1087, scale=0.55]{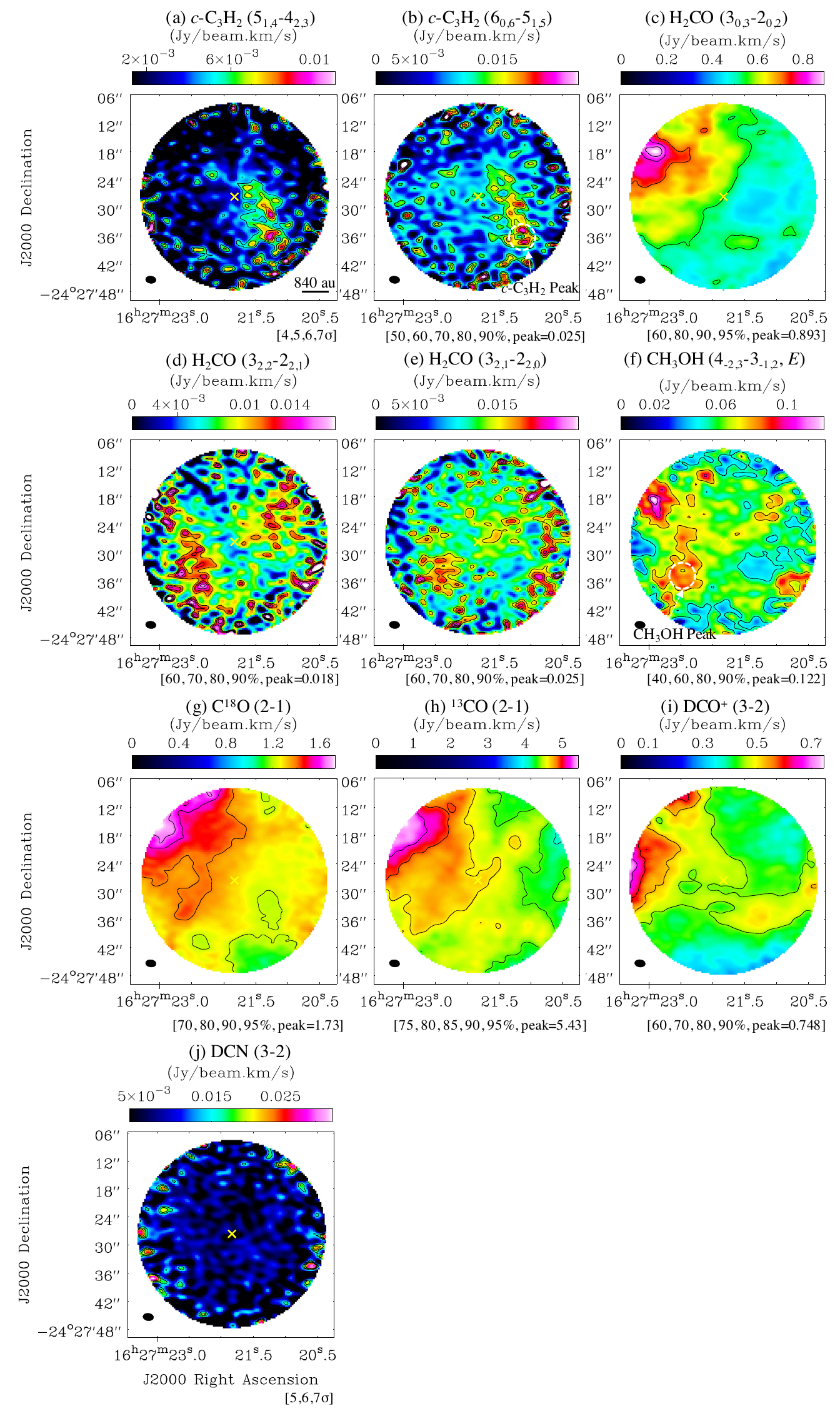}
 \end{center}
\caption{Moment 0 maps (combination of 12-m array, 7-m array, and TP) of $c$-C$_{3}$H$_{2}$ in panels (a) and (b), H$_{2}$CO in panels (c)--(e), CH$_{3}$OH in panel (f), C$^{18}$O in panel (g), $^{13}$CO in panel (h), DCO$^{+}$ in panel (i), and DCN in panel (j) toward Oph-emb5. The yellow crosses indicate positions of the infrared sources \citep{2009ApJ...692..973E}. The contour levels are indicated in each panel at the bottom. The use of ``$\sigma$'' means that contours are based on the rms noise level, and using ``\%'' means contours relative to the peak intensity. The rms noise levels for each panel are summarized in Table \ref{tab:mom0}. The filled black ellipses indicate the beam sizes of 2\farcs2 $\times$ 1\farcs5. In panels (b) and (f), $c$-C$_{3}$H$_{2}$ Peak and CH$_{3}$OH Peak are indicated as white dashed circles (see Section \ref{sec:specana}). \label{fig:EES2009mom0}}
\end{figure*}

\begin{figure*}[!th]
\figurenum{4}
 \begin{center}
  \includegraphics[bb = 0 30 663 1082, scale=0.55]{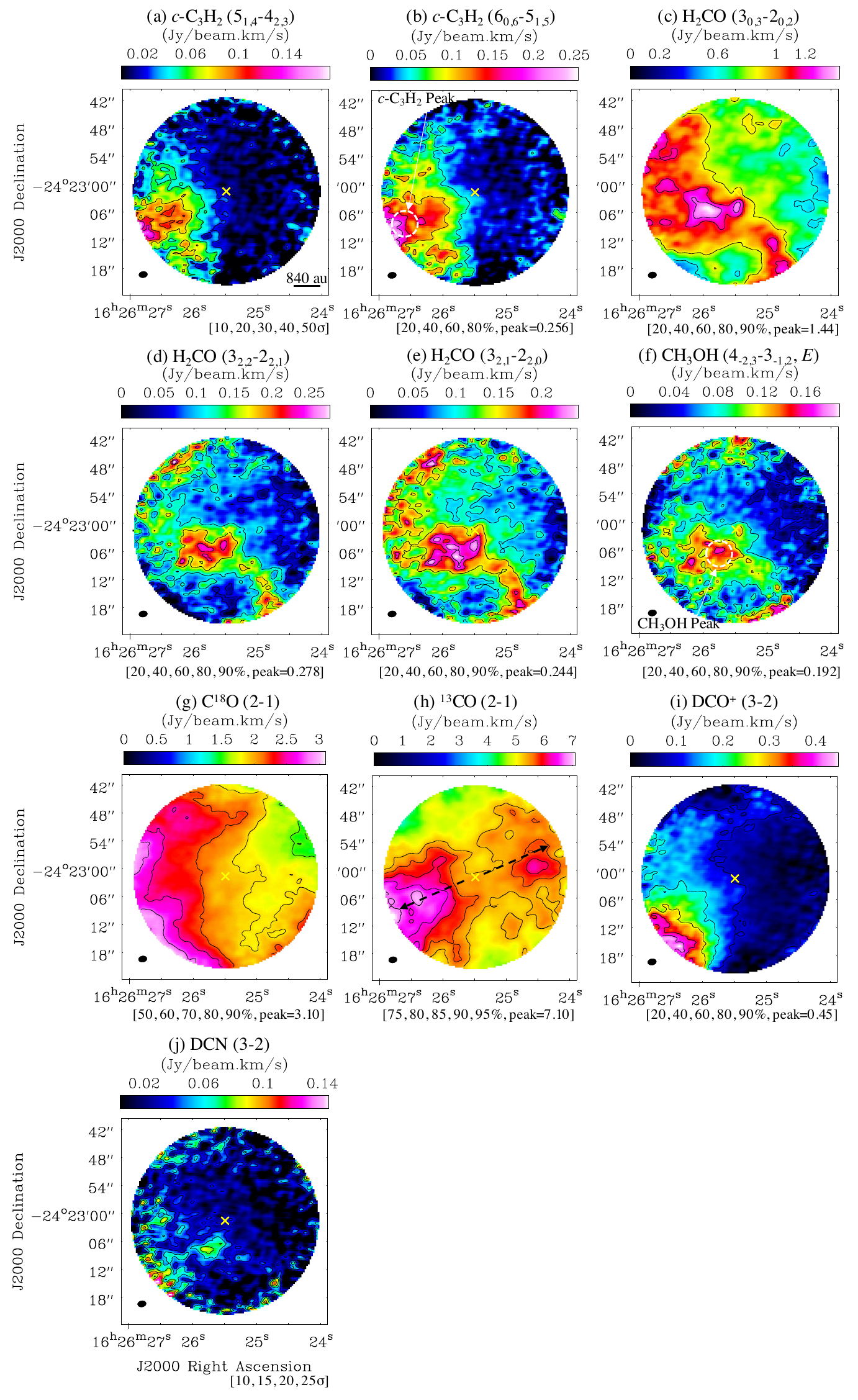}
 \end{center}
\caption{Moment 0 maps (combination of 12-m array, 7-m array, and TP) of $c$-C$_{3}$H$_{2}$ in panels (a) and (b), H$_{2}$CO in panels (c)--(e), CH$_{3}$OH in panel (f), C$^{18}$O in panel (g), $^{13}$CO in panel (h), DCO$^{+}$ in panel (i), and DCN in panel (j) toward Oph-emb9. The yellow crosses indicate positions of the infrared sources \citep{2009ApJ...692..973E}. The contour levels are indicated in each panel at the bottom. The use of ``$\sigma$'' means that contours are based on the rms noise level, and using ``\%'' means contours relative to the peak intensity. The rms noise levels for each panel are summarized in Table \ref{tab:mom0}. The filled black ellipses indicate the beam sizes of 1\farcs8 $\times$ 1\farcs3. In panels (b) and (f), $c$-C$_{3}$H$_{2}$ Peak and CH$_{3}$OH Peak are indicated as white dashed circles (see Section \ref{sec:specana}). The black dashed arrows in panel (h) indicate the direction of the molecular outflow \citep{2019A&A...626A..71A}. \label{fig:GY92mom0}}
\end{figure*}

%Table3
\begin{deluxetable*}{cclcccc}
\tablenum{3}
\tablecaption{Summary of the moment 0 maps\label{tab:mom0}}
\tablewidth{0pt}
\tablehead{
\colhead{Panel} & \colhead{Species} & \colhead{Transition} & \colhead{Rest Frequency} & \colhead{$E_{\rm {up}}/k$} & \multicolumn{2}{c}{rms\tablenotemark{a}} \\
\cline{6-7}
\colhead{} & \colhead{} & \colhead{} & \colhead{(GHz)} & \colhead{(K)} & \colhead{Oph-emb5} & \colhead{Oph-emb9}
}
%\decimalcolnumbers
\startdata
(a) & $c$-C$_{3}$H$_{2}$ & $5_{1, 4}- 4_{2, 3}$ & 217.940046 & 35.4 & 1.3 & 2.5 \\
(b) & $c$-C$_{3}$H$_{2}$ & $6_{0, 6}- 5_{1, 5}$ & 217.822148 & 38.6  & 1.0 & 3.2 \\
(c) & H$_{2}$CO & $3_{0, 3}-2_{0, 2}$ & 218.222192 & 21.0  & 2.4 & 6.9 \\
(d) & H$_{2}$CO & $3_{2, 2}- 2_{2, 1}$ & 218.475632 & 68.1 & 1.8 & 6.1 \\
(e) & H$_{2}$CO & $3_{2, 1}- 2_{2, 0}$ & 218.760066 & 68.1 & 1.3 & 5.5 \\
(f) & CH$_{3}$OH & $4_{-2, 3}- 3_{-1, 2}$ $E$ & 218.440063 & 45.5 & 1.5 & 5.1 \\
(g) & C$^{18}$O & $2-1$ & 219.5603541 & 15.8 & 3.5 & 6.4 \\
(h) & $^{13}$CO & $2-1$ & 220.3986842 & 15.9 & 5.6 & 10.9 \\ 
(i) & DCO$^{+}$ & $3-2$ & 216.1125822 & 20.7 & 1.5 & 3.4 \\ 
(j) & DCN & $3-2$ & 217.2385378 & 20.9 & 3.1 & 3.8 \\
\enddata
\tablenotetext{a}{Unit is mJy\,beam$^{-1}$\,km\,s$^{-1}$.}
\tablecomments{Transition, rest frequency, and upper-state energy are taken from the Cologne Database for Molecular Spectroscopy \citep[CDMS;][]{2005JMoSt.742..215M}.}
\end{deluxetable*}

Figures \ref{fig:EES2009mom0} and \ref{fig:GY92mom0} show moment 0 maps of the observed molecular lines toward Oph-emb5 and Oph-emb9, respectively.
Table \ref{tab:mom0} summarizes information on the noise levels for each panel.

Around Oph-emb5, there are differences in spatial distributions between carbon-chain species ($c$-C$_{3}$H$_{2}$) and COMs (H$_{2}$CO and CH$_{3}$OH).
The hydrocarbon $c$-C$_{3}$H$_{2}$ is enhanced in the region west of the IR source (shown in panels (a) and (b) in Figure \ref{fig:EES2009mom0}).
On the other hand, H$_{2}$CO and CH$_{3}$OH tend to be enhanced to the east of the IR source.
The H$_{2}$CO ($3_{0, 3}-2_{0, 2}$) line (panel (c)) is more extended than its other lines (panels (d) and (e)).
This seems to be caused by the different upper-state energies (Table \ref{tab:mom0}).
The peaks of CH$_{3}$OH are consistent with those of H$_{2}$CO.
Peaks of C$^{18}$O and $^{13}$CO (panels (g) and (h)) are located at the north-east edge and their spatial distributions resemble that of H$_{2}$CO ($3_{0,3}-2_{0,2}$) line.
The DCO$^{+}$ moment 0 map shows an extended structure (panel (i)).
We did not detect the DCN line toward Oph-emb5 (panel (j)).

In Oph-emb9, the observed molecular emission for all detected lines is enhanced in the eastern part of the FoV.
This side is irradiated by the nearby Herbig Be star (Figure \ref{fig:wise}). More details are discussed in Section \ref{sec:d1}.
H$_{2}$CO shows an elongated feature from the northeast to the southwest. 
The CH$_{3}$OH emission shows a similar tendency as H$_{2}$CO, and their peaks are consistent with each other.
In addition, a weak CH$_{3}$OH peak is associated with the IR source.
The emission from C$^{18}$O and $^{13}$CO is generally strong in the eastern edge of the FoV, and $^{13}$CO shows an additional, less strong and more concentrated, emission peak to the west.
The orientation of the $^{13}$CO emission peaks is consistent with the molecular outflow \citep{2019A&A...626A..71A}.
We have detected the DCN line toward Oph-emb9.
Its spatial distribution is the most compact and is different from the other deuterated species, DCO$^{+}$.
The upper state energy of the observed DCN line ($E_{\rm {up}}/k=20.9$ K) is similar to that of DCO$^{+}$ (20.7 K), and it follows that the different spatial distributions are not caused by different excitation conditions.
The details are discussed in Section \ref{sec:d4}.

\subsubsection{Comparison of spatial distributions of $c$-C$_{3}$H$_{2}$ with H$_{2}$CO and CH$_{3}$OH} \label{sec:d1}

In this subsection, we investigate effects of the UV radiation from the Herbig Be star on the chemistry around the observed YSOs by comparing the spatial distributions of $c$-C$_{3}$H$_{2}$ with H$_{2}$CO and CH$_{3}$OH.
In astrochemical models, $c$-C$_{3}$H$_{2}$ can be formed by the destruction of large hydrocarbons through UV irradiation (top-down chemistry) or through the accumulation of C and H atoms to form small hydrocarbons (bottom-up chemistry) \citep{2018A&A...617A.120M}.
Large hydrocarbons could also be formed by destruction of dust grains \citep[e.g.,][]{2014ApJ...797L..30Z}.
In a more specific bottom-up chemistry, small hydrocarbons can be efficiently formed in the gas phase via ion-molecule reactions including C$^{+}$ \citep[e.g.,][]{1993ApJ...417..181M, 2017A&A...605A..88L}.

In the case of Oph-emb5, $c$-C$_{3}$H$_{2}$ is enhanced at the west side, whereas  H$_{2}$CO and CH$_{3}$OH are enhanced at the eastern parts relative to the YSO.
Since Oph-emb5 is unlikely affected by any sources \citep{2017ApJ...835....3L}, these features are not induced by the UV radiation, but may indicate effects of stellar feedback, such as the molecular outflows.
We discuss this point further in Section \ref{sec:d3}.

In Oph-emb9, the spatial distributions of $c$-C$_{3}$H$_{2}$ show peaks at the eastern edge of the field, which is irradiated by the Herbig Be star.
On the other hand, the spatial distributions of H$_{2}$CO and CH$_{3}$OH show maxima that are located farther from the Herbig Be star, and closer to the center of the field.
The differences of the spatial distributions between small hydrocarbons and COMs in Oph-emb9 imply that top-down chemistry may efficiently contribute to the $c$-C$_{3}$H$_{2}$ formation under conditions of UV irradiation.

\subsubsection{Comparison of spatial distributions of DCO$^{+}$ and DCN} \label{sec:d4}

In this subsection, we discuss the higher DCO$^{+}$ column density in Oph-emb9 compared with Oph-emb5 and the sole DCN detection in Oph-emb9 as seen in their moment 0 maps.

The DCO$^{+}$ ion is considered to be formed in the gas phase mainly by the following reaction in cold environments \citep[$< 30$ K;][]{2014prpl.conf..859C}:
\begin{equation} \label{equ:DCO1}
{\rm {H}}_{2}{\rm {D}}^{+} + {\rm {CO}} \rightarrow   {\rm {DCO}}^{+} + {\rm {H}}_{2}.
\end{equation}
Another reaction that could produce DCO$^{+}$ in warm regions \citep[$>30$ K;][]{1985ApJ...294L..63A} is:
\begin{equation} \label{equ:DCO2}
{\rm {HCO}}^{+} + {\rm {D}} \rightarrow  {\rm {DCO}}^{+} + {\rm {H}}.
\end{equation}

In the case of DCN, the main formation pathways \citep{1989ApJ...340..906M,2001ApJS..136..579T} are:
\begin{equation} \label{equ:DCN1}
{\rm {CH}}_{2}{\rm {D}}^{+} + {\rm {H}}_{2} \rightarrow {\rm {CH}}_{4}{\rm {D}}^{+} + h\nu,
\end{equation}
\begin{equation} \label{equ:DCN2}
{\rm {CH}}_{4}{\rm {D}}^{+} + {\rm {e}}^{-} \rightarrow {\rm {CHD}} + {\rm {H}}_{2} + {\rm {H}},
\end{equation}
followed by
\begin{equation} \label{equ:DCN3}
{\rm {CHD}} + {\rm {N}} \rightarrow {\rm {DCN}} + {\rm {H}},
\end{equation}
or,
\begin{equation} \label{equ:DCN4}
{\rm {CH}}_{2}{\rm {D}}^{+} + {\rm {e}}^{-} \rightarrow {\rm {CHD}}+ {\rm {H}},
\end{equation}
followed by Reaction (\ref{equ:DCN3}).
Hence, DCN is mainly formed by CH$_{2}$D$^{+}$.
The CH$_{2}$D$^{+}$ ion is formed by reaction between CH$_{3}^{+}$ and HD, and the endothermicity of the back reaction is 654 K, which is higher than the reaction between H$_{3}^{+}$ and HD (232 K) which forms H$_{2}$D$^{+}$ \citep{2021PhR...893....1O}. 
As a result, CH$_{2}$D$^{+}$ survives more easily in warmer gas ($30<T<100$ K) compared to H$_{2}$D$^{+}$ \citep[$<30$ K;][]{2014prpl.conf..859C}.
This means that DCN can efficiently form in warmer regions, whereas DCO$^{+}$ can exist both in cold and warm regions due to  Reactions (\ref{equ:DCO1}) and  (\ref{equ:DCO2}).

In Oph-emb9, the UV radiation from the Herbig Be star can heat the gas and produce both the precursor ions HCO$^{+}$ and  CH$_{2}$D$^{+}$ \citep{2017ApJ...835....3L}, leading to DCO$^{+}$ (via Reaction (\ref{equ:DCO2}), $>30$ K) and DCN (via Reactions (\ref{equ:DCN1}) to (\ref{equ:DCN4}), $>30$ K). 
In fact, the DCN spatial distribution follows that of the irradiated material, where the gas temperature is expected to be higher (Figure \ref{fig:GY92mom0}). 
On the other hand, in the case of Oph-emb5, the non-detection of DCN results from lower-temperature conditions, which implies formation of DCO$^{+}$ via Reaction (\ref{equ:DCO1}) ($< 30$ K). 
The lower column density of  DCO$^{+}$ in  Oph-emb5 compared to Oph-emb9 corresponds to lower H$_{2}$D$^{+}$ abundance due to weaker UV irradiation.

\subsection{Spectral analysis} \label{sec:specana}

The spectral analysis of the detected lines of H$_{2}$CO, CH$_{3}$OH, $c$-C$_{3}$H$_{2}$, DCO$^{+}$, and DCN was carried out using the CASSIS software \citep{2015sf2a.conf..313V} together with the CDMS\footnote{\url{https://cdms.astro.uni-koeln.de}} and JPL\footnote{\url{https://spec.jpl.nasa.gov}} spectroscopic databases.
For the spectral analysis, we constructed 6\arcsec (840 au) beam average spectra to balance the angular resolution and sensitivity. 
We pick two positions around each YSO, which are the strong $c$-C$_{3}$H$_{2}$ emission positions, and the strong CH$_{3}$OH emission positions. 
They are chosen because each position seems to represent different chemical features.
We indicate these locations as ``$c$-C$_{3}$H$_{2}$ Peak'' and ``CH$_{3}$OH Peak'', respectively.
Panels (b) and (f) in Figures \ref{fig:EES2009mom0} and \ref{fig:GY92mom0} indicate the positions and the beam (6\arcsec) used for spectral analysis.

We applied the Markov chain Monte Carlo (MCMC) method assuming the local thermodynamic equilibrium (LTE) model in CASSIS.
In this method, the column density ($N$), excitation temperature ($T_{\rm {ex}}$), line width (FWHM), and radial velocity ($V_{\rm {LSR}}$) were treated as semi-free parameters within certain ranges, and solutions were obtained by a $\chi^2$ minimization.
The excitation temperatures of $c$-C$_{3}$H$_{2}$ and H$_{2}$CO were derived to be $8.7\pm1.0$ K and $15.5\pm0.7$ K in Oph-emb5, $9.9\pm0.3$ K and $36.2\pm0.4$ K in Oph-emb9, respectively, based on APEX observations \citep{2017ApJ...835....3L}.
We then set the excitation temperature range from 10 K to 40 K.

Table \ref{tab:CASSIS} summarizes fitting results. Since the H$_{2}$CO ($3_{0, 3}-2_{0, 2}$) line has different spatial distributions (Figures \ref{fig:EES2009mom0} and \ref{fig:GY92mom0}) and different upper state energies (Table \ref{tab:mom0}) from the other two lines, these lines probably do not trace the same regions.
We then derived parameters for H$_{2}$CO with the following two cases; (a) using the $3_{0, 3}-2_{0, 2}$ line, and (b) using the $3_{2, 2}- 2_{2, 1}$ and $3_{2, 1}- 2_{2, 0}$ lines.

\begin{figure*}[!th]
\figurenum{5}
 \begin{center}
  \includegraphics[bb = 9 10 536 670, scale=0.8]{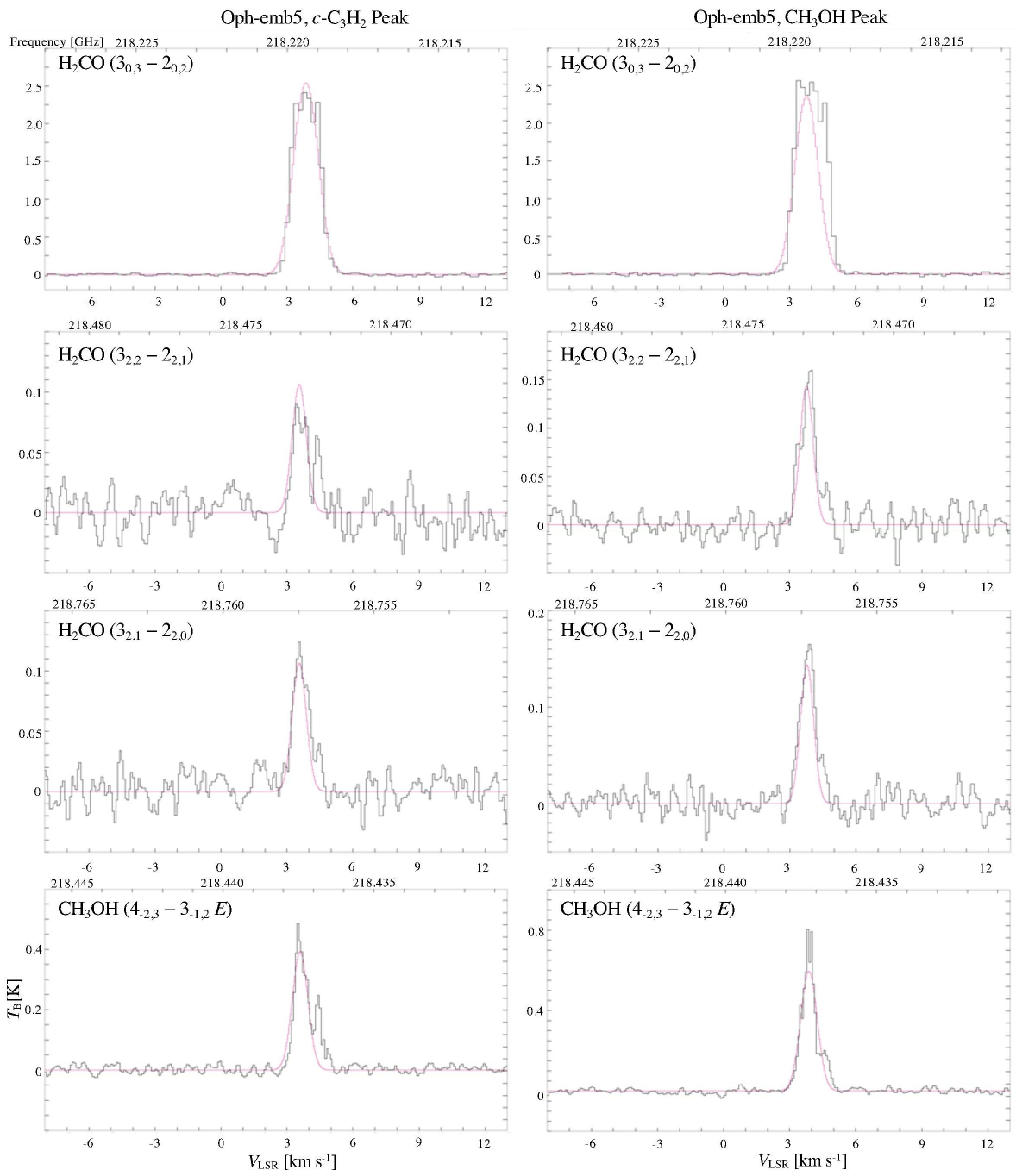}
 \end{center}
\caption{Spectra of H$_{2}$CO and CH$_{3}$OH toward Oph-emb5. Purple curves indicate the fitted results obtained using the CASSIS software. \label{fig:EES2009_spec1}}
\end{figure*}

\begin{figure*}[!th]
\figurenum{6}
 \begin{center}
  \includegraphics[bb = 3 10 579 500, scale=0.8]{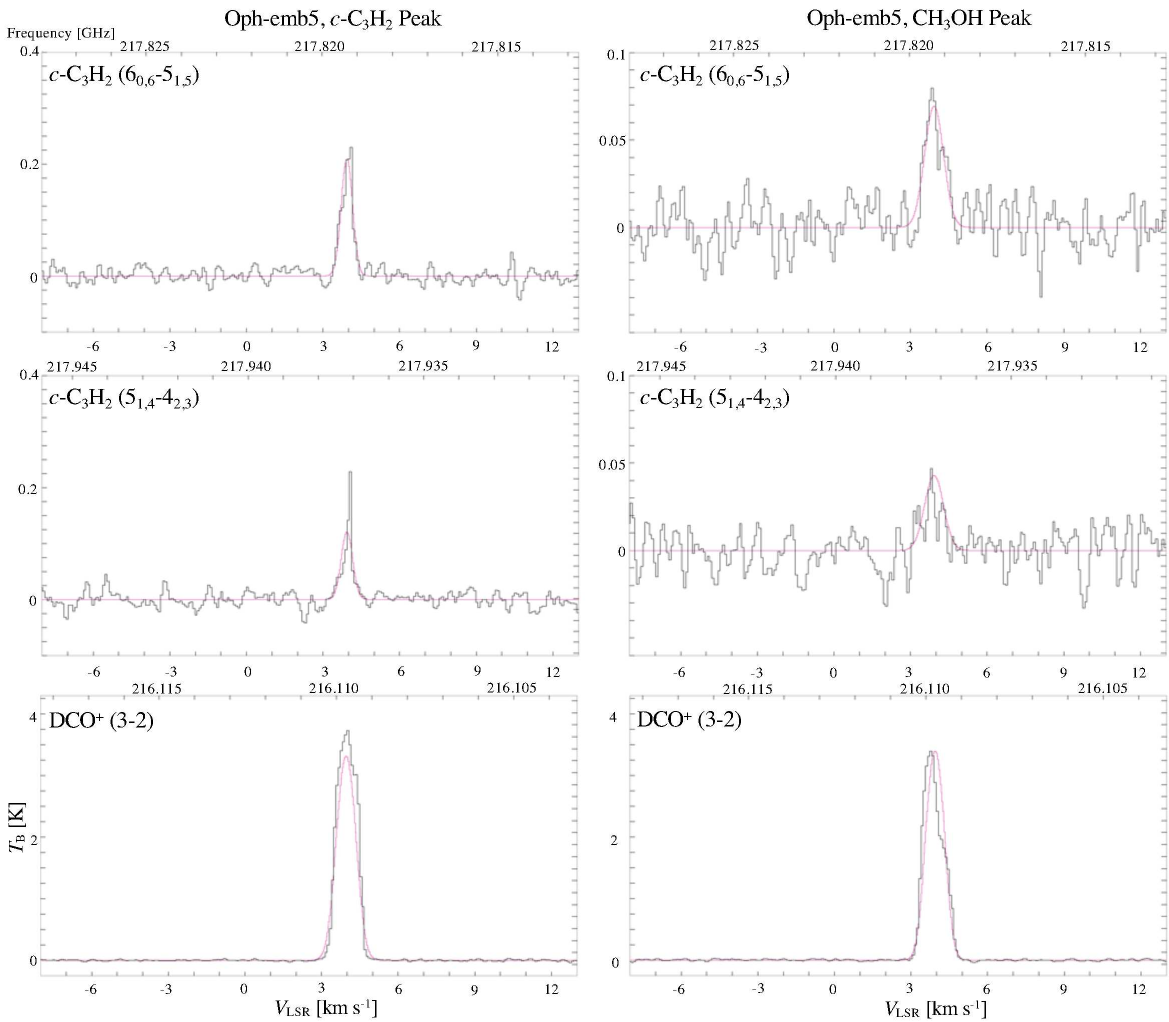}
 \end{center}
\caption{Spectra of $c$-C$_{3}$H$_{2}$ and DCO$^{+}$ toward Oph-emb5. Purple curves indicate the fitted results obtained using the CASSIS software. \label{fig:EES2009_spec2}}
\end{figure*}

\begin{figure*}[!th]
\figurenum{7}
 \begin{center}
  \includegraphics[bb = 3 10 532 650, scale=0.8]{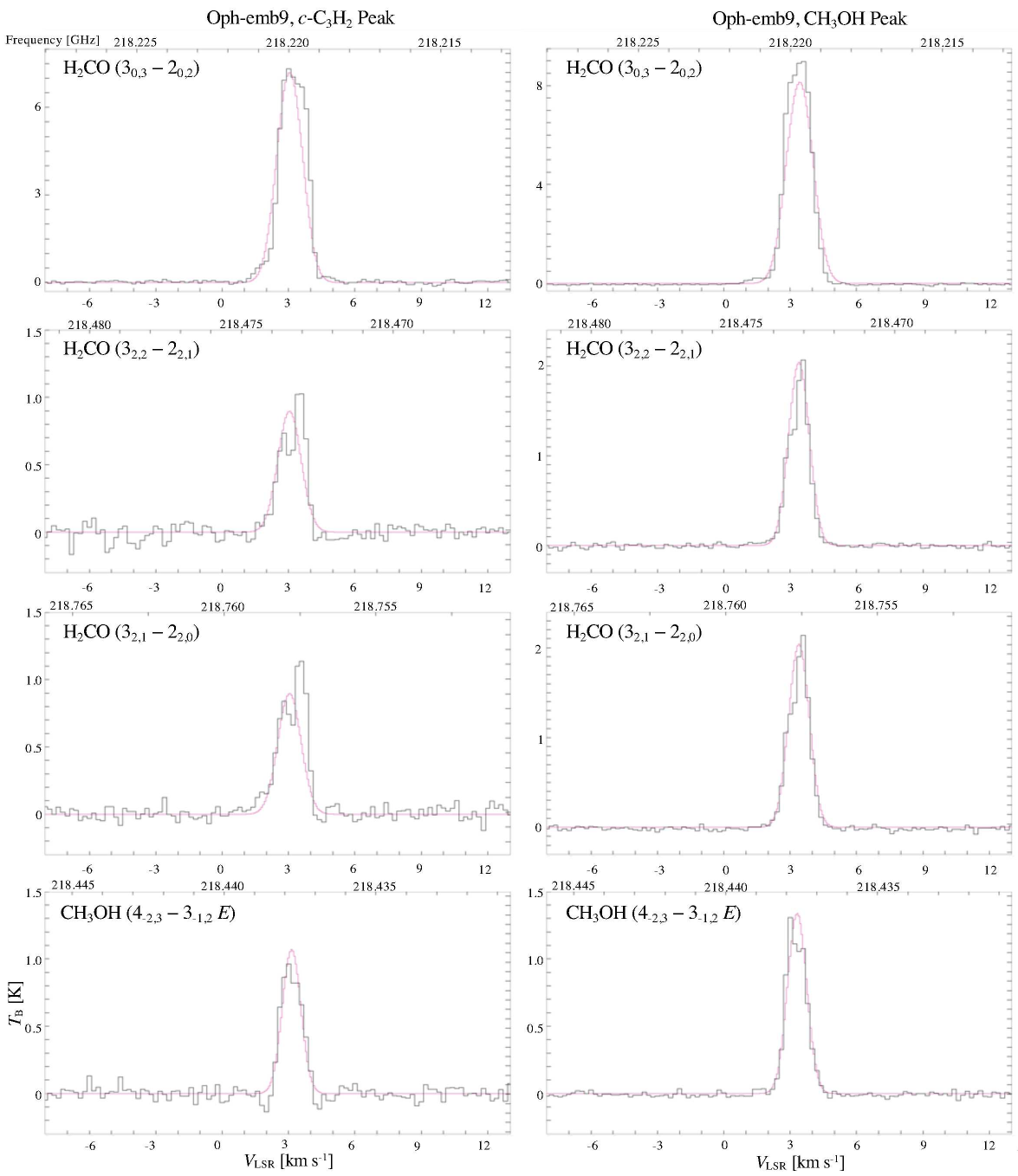}
 \end{center}
\caption{Spectra toward of H$_{2}$CO and CH$_{3}$OH Oph-emb9. Purple curves indicate the fitted results obtained using the CASSIS software. \label{fig:GY92_spec1}}
\end{figure*}

\begin{figure*}[!th]
\figurenum{8}
 \begin{center}
  \includegraphics[bb = 4 10 537 660, scale=0.8]{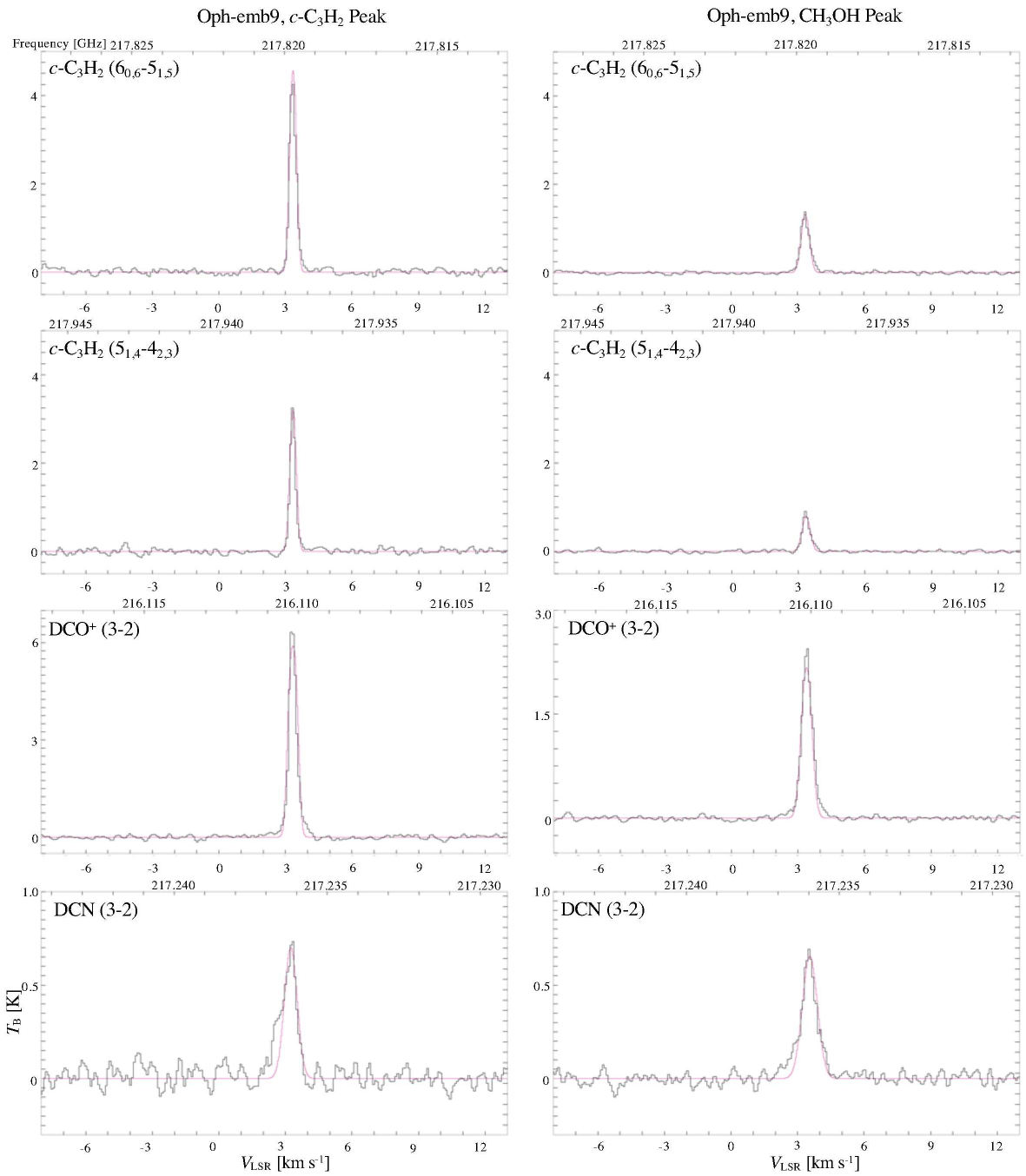}
 \end{center}
\caption{Spectra $c$-C$_{3}$H$_{2}$, DCO$^{+}$, and DCN toward Oph-emb9. Purple curves indicate the fitted results obtained using the CASSIS software. \label{fig:GY92_spec2}}
\end{figure*}

Figures \ref{fig:EES2009_spec1} and \ref{fig:EES2009_spec2} show spectra at the $c$-C$_{3}$H$_{2}$ peak and CH$_{3}$OH peak in Oph-emb5, and Figures \ref{fig:GY92_spec1} and \ref{fig:GY92_spec2} show spectra at each position in Oph-emb9.
Purple curves indicate the fitted results (Table \ref{tab:CASSIS}). 
The second velocity components may have been detected in the H$_{2}$CO and CH$_{3}$OH spectra, especially at the $c$-C$_{3}$H$_{2}$ peak of Oph-emb5. 
However, we cannot fit these second components well with the current velocity resolution, and we therefore did not consider the second velocity component in our analyses.
The radial velocities are almost consistent with previous APEX observations \citep{2017ApJ...835....3L}.

\floattable
\rotate
\begin{deluxetable*}{cccccccccc}
\tabletypesize{\scriptsize}
\tablenum{4}
\tablecaption{Summary of analytical results of spectra by CASSIS\label{tab:CASSIS}}
\tablewidth{0pt}
\tablehead{
                 \colhead{}& \multicolumn{4}{c}{$c$-C$_{3}$H$_{2}$ Peak} & & \multicolumn{4}{c}{CH$_{3}$OH Peak} \\
		       \cline{2-5} \cline{7-10} 
		    \colhead{Species}  & \colhead{$N$ (cm$^{-2}$)} & \colhead{$T_{\rm {ex}}$ (K)} & \colhead{FWHM (km\,s$^{-1}$)} & \colhead{$V_{\rm {LSR}}$ (km\,s$^{-1}$)} & & \colhead{$N$ (cm$^{-2}$)} & \colhead{$T_{\rm {ex}}$ (K)} & \colhead{FWHM (km\,s$^{-1}$)} & \colhead{$V_{\rm {LSR}}$ (km\,s$^{-1}$)}
}
\startdata
\multicolumn{10}{c}{{\bf {Oph-emb5}}} \\
	$c$-C$_{3}$H$_{2}$ & ($7.7 \pm 1.2$)$\times 10^{11}$ & $15.3 \pm 2.2$ & $0.57 \pm 0.09$ & $3.94 \pm 0.04$ & & ($6.3 \pm 1.0$)$\times 10^{11}$ & $12.0 \pm1.4$ & $0.83 \pm 0.11$ & $3.93 \pm 0.11$ \\
	H$_{2}$CO\tablenotemark{a} & ($2.9 \pm 2.1$)$\times 10^{13}$ & $11.6 \pm 0.9$ & $1.36 \pm 0.09$ & $3.78 \pm 0.09$ & & ($1.7 \pm 0.5$)$\times 10^{13}$ & $16.7 \pm 1.9$ & $1.28 \pm 0.07$ & $3.763 \pm 0.005$ \\
	H$_{2}$CO\tablenotemark{b} & ($5.6 \pm 4.3$)$\times 10^{13}$ & $14.0 \pm 3.6$ & $0.78 \pm 0.08$ & $3.52 \pm 0.08$ & & ($1.52 \pm 0.08$)$\times 10^{13}$ & $17.8 \pm 0.9$ & $0.63 \pm 0.07$ & $3.765 \pm 0.004$ \\
	CH$_{3}$OH & ($5.3 \pm 0.9$)$\times 10^{13}$ & $29.9 \pm 4.7$ & $0.84 \pm 0.07$ & $3.53 \pm 0.08$ & & ($1.1 \pm 0.5$)$\times 10^{14}$ & $21.0 \pm 7.0$ & $0.82 \pm 0.07$ & $3.87 \pm 0.02$ \\
	DCO$^{+}$ & ($2.5 \pm 0.2$)$\times 10^{12}$ & $11.7 \pm 0.5$ & $0.83 \pm 0.02$ & $3.98 \pm 0.03$ & & ($4.4 \pm 1.1$)$\times 10^{12}$ & $11.3 \pm 0.6$ & $0.73 \pm 0.07$ & $3.70 \pm 0.18$ \\
	\cline{2-5} \cline{7-10} 
	$N$($c$-C$_{3}$H$_{2}$)/$N$(CH$_{3}$OH)\tablenotemark{c} & $0.015 \pm 0.003$ & \multicolumn{3}{l}{} & & $0.006 \pm 0.003$ & \multicolumn{3}{l}{} \\
%	$N$(H$_{2}$CO)/$N$(CH$_{3}$OH)\tablenotemark{d} & $1.0 \pm 0.8$ & \multicolumn{3}{l}{} & & $0.13 \pm 0.06$ & \multicolumn{3}{l}{} \\
	\cline{1-10} 
\multicolumn{10}{c}{{\bf {Oph-emb9}}} \\
	$c$-C$_{3}$H$_{2}$ & ($6.7 \pm 1.0$)$\times10^{13}$ & $10.5 \pm 0.3$ & $0.59 \pm 0.05$ & $3.31 \pm 0.10$ & & ($4.7 \pm 2.5$)$\times10^{12}$ & $12.3 \pm 1.9$ & $0.43 \pm 0.04$ & $3.30 \pm 0.09$ \\
	H$_{2}$CO\tablenotemark{a} & ($1.24 \pm 0.06$)$\times10^{14}$ & $38.3 \pm 0.7$ & $1.31 \pm 0.01$ & $3.12 \pm 0.05$ & & ($1.54 \pm 0.06$)$\times10^{14}$ & $33.7 \pm 1.8$ & $1.40 \pm 0.04$ & $3.28 \pm 0.10$ \\
	H$_{2}$CO\tablenotemark{b} & ($8.1 \pm 0.9$)$\times10^{13}$ & $38.3 \pm 1.2$ & $1.36 \pm 0.10$ & $3.19 \pm 0.12$ & & ($2.2 \pm 0.5$)$\times 10^{14}$ & $26.2 \pm 5.3$ & $1.03 \pm 0.09$ & $3.39 \pm 0.05$ \\
	CH$_{3}$OH & ($1.8 \pm 0.1$)$\times 10^{14}$ & $35.9 \pm 2.6$ & $0.97 \pm 0.04$ & $3.15 \pm 0.03$ & & ($2.4 \pm 0.2$)$\times 10^{14}$ & $34.4 \pm 3.1$ & $1.02 \pm 0.05$ & $3.28 \pm 0.04$ \\
	DCO$^{+}$ & ($1.1 \pm 0.8$)$\times 10^{13}$ & $11.2 \pm 0.5$ & $0.40 \pm 0.08$ & $3.26 \pm 0.11$ & & ($9.3 \pm 3.2$)$\times 10^{11}$ & $11.0 \pm 0.5$ & $0.56 \pm 0.04$ & $3.40 \pm 0.06$ \\
	DCN & ($9.4 \pm 1.7$)$\times 10^{11}$ & $10.5 \pm 0.2$ & $0.69 \pm 0.03$ & $3.23 \pm 0.01$ & & ($7.3 \pm 0.7$)$\times10^{11}$ & $10.5 \pm 0.2$ & $0.85 \pm 0.06$ & $3.58 \pm 0.03$ \\
	\cline{2-5} \cline{7-10} 
	$N$($c$-C$_{3}$H$_{2}$)/$N$(CH$_{3}$OH)\tablenotemark{c} & $0.37 \pm 0.06$ & \multicolumn{3}{l}{} & & $0.020 \pm 0.011$ & \multicolumn{3}{l}{} \\
%	$N$(H$_{2}$CO)/$N$(CH$_{3}$OH)\tablenotemark{d} & $0.44 \pm 0.06$ & \multicolumn{3}{l}{} & & $0.9 \pm 0.2$ & \multicolumn{3}{l}{} \\
\enddata
		\tablecomments{The errors are the standard deviation.}
		\tablenotetext{a}{The values are derived from fitting of the $3_{0, 3}-2_{0, 2}$ line.}
		\tablenotetext{b}{The values are derived from fitting of the $3_{2, 2}- 2_{2, 1}$ and $3_{2, 1}- 2_{2, 0}$ lines.}
		\tablenotetext{c}{The errors are calculated from the standard deviation of the column densities.}
		%\tablenotetext{d}{The values are calculated using case ``b'' of $N$(H$_{2}$CO). The errors are calculated from the standard deviation of the column densities.}
\end{deluxetable*}

\subsubsection{Comparison of the $N$($c$-C$_{3}$H$_{2}$)/$N$(CH$_{3}$OH) ratio derived from CASSIS} \label{sec:d2}

UV radiation, if sufficiently intense, produces photon-dominated regions (PDRs).
In PDR chemistry, small hydrocarbons such as $c$-C$_{3}$H$_{2}$ can be abundant in less shielded regions \citep[e.g., $A_{\rm {v}} \sim 1.43$ mag;][]{2019A&A...626A..28T}, while CH$_{3}$OH formation is efficient in more shielded regions \citep[$A_{\rm {v}}>2$ mag;][]{2019MNRAS.486.1853E}.
In order to investigate the effects of UV irradiation from the nearby Herbig Be star on Oph-emb9, we derive the $N$($c$-C$_{3}$H$_{2}$)/$N$(CH$_{3}$OH) ratio at each position as summarized in Table \ref{tab:CASSIS}. The ratios at the $c$-C$_{3}$H$_{2}$ Peak and the CH$_{3}$OH Peak in Oph-emb5 are derived to be $0.015\pm0.003$ ($1\sigma$) and $0.006\pm0.003$, and those in Oph-emb9 are calculated as $0.37\pm0.06$ and $0.020\pm0.011$, respectively.

As a general trend, the $N$($c$-C$_{3}$H$_{2}$)/$N$(CH$_{3}$OH) ratios around Oph-emb9 are higher than those around Oph-emb5.
This means that $c$-C$_{3}$H$_{2}$ is relatively more abundant compared to CH$_{3}$OH in Oph-emb9. 
Hence, this region may be affected by the UV radiation from the Herbig Be star.

We compare these ratios at the $c$-C$_{3}$H$_{2}$ Peak and the CH$_{3}$OH Peak around each YSO.
In the case of Oph-emb5, the $N$($c$-C$_{3}$H$_{2}$)/$N$(CH$_{3}$OH) ratio at the $c$-C$_{3}$H$_{2}$ Peak is higher than that at the CH$_{3}$OH Peak by a factor of $\sim2.6$.
In Oph-emb9, the difference between the $c$-C$_{3}$H$_{2}$ Peak and the CH$_{3}$OH Peak is a factor of $\sim18.5$.
Hence, $c$-C$_{3}$H$_{2}$ is significantly enhanced at the $c$-C$_{3}$H$_{2}$ Peak in Oph-emb9 compared to Oph-emb5.
All of these results also support the PDR chemistry around Oph-emb9, which is likely driven by the nearby Herbig Be star.
The PDR chemistry in this source was also suggested by the layered structures of CO and its isotopologues \citep{2019ApJ...875...62Y}.
Our conclusion that the chemistry in the field of view of Oph-emb9 is strongly affected by the Herbig Be star is consistent with their results.

\subsection{Moment 2 Maps of the $^{13}$CO and C$^{18}$O lines} \label{sec:resmom2}

\begin{figure*}[!th]
\figurenum{9}
 \begin{center}
  \includegraphics[bb = 0 20 454 449, scale=0.6]{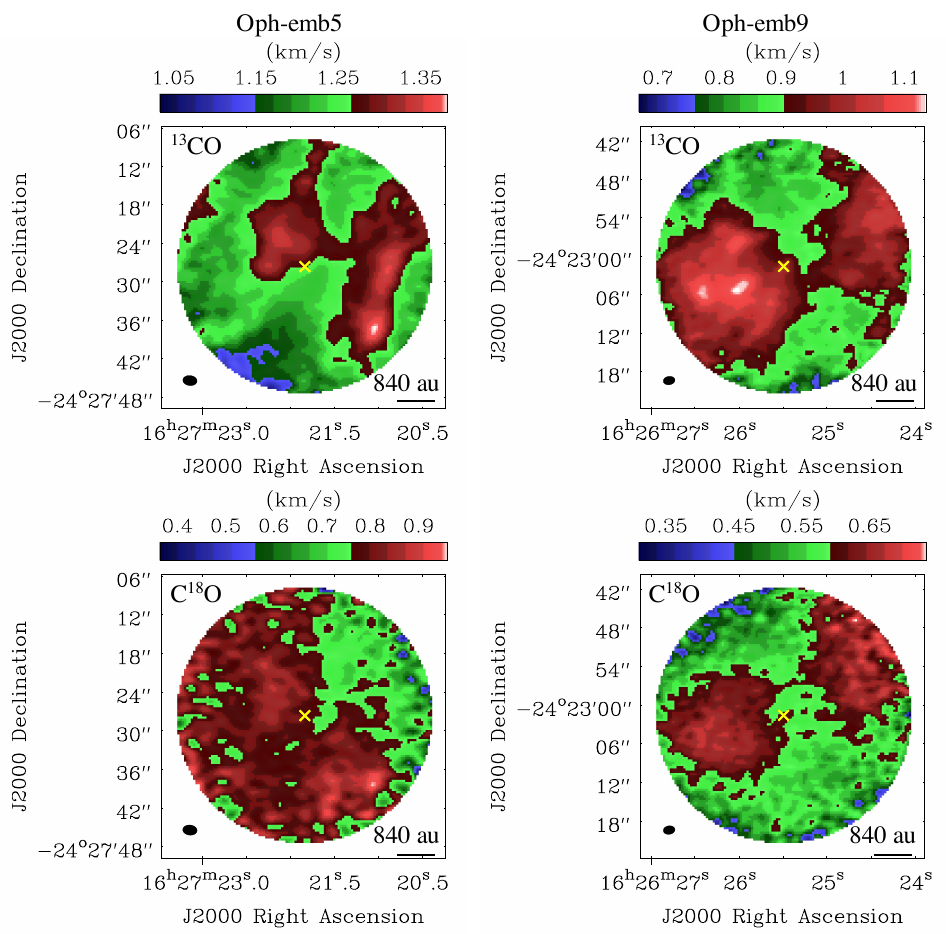}
 \end{center}
\caption{Moment 2 maps of the $^{13}$CO ($J=2-1$) and C$^{18}$O ($J=2-1$) lines in upper panels and bottom panels, respectively. Left and right panels show results for Oph-emb5 and Oph-emb9, respectively. The velocity ranges used in the $^{13}$CO and C$^{18}$O maps are -1.2 -- +6 km\,s$^{-1}$ and -0.6 -- +5.8 km\,s$^{-1}$ in Oph-emb5, and 0 -- +6.4 km\,s$^{-1}$ and +0.8 -- +5.6 km\,s$^{-1}$ in Oph-emb9, respectively.The yellow crosses indicate positions of the infrared sources \citep{2009ApJ...692..973E}. The filled black ellipses indicate the angular resolutions of 2\farcs2 $\times$ 1\farcs5 and 1\farcs8 $\times$ 1\farcs3 for Oph-emb5 and Oph-emb9, respectively. \label{fig:mom2}}
\end{figure*}

Figure \ref{fig:mom2}  shows moment 2 maps (velocity dispersion maps) of the $^{13}$CO ($J=2-1$) and C$^{18}$O ($J=2-1$) lines toward Oph-emb5 and Oph-emb9, respectively.
In Appendix \ref{sec:appendix1}, channel maps of these lines are presented in Figures \ref{fig:EES2009_13CO_channel}--\ref{fig:GY92_C18O_channel}. Around Oph-emb5, there are two velocity 
dispersion peaks in the $^{13}$CO moment 2 maps. The moment 2 maps of $^{13}$CO and C$^{18}$O are similar to each other around Oph-emb9; two dispersion maxima are located at the eastern and western positions, respectively.

\subsubsection{Comparisons of molecular distributions and velocity dispersion} \label{sec:d3}

\begin{figure*}[!th]
\figurenum{10}
 \begin{center}
  \includegraphics[bb = 0 20 444 445, scale=0.6]{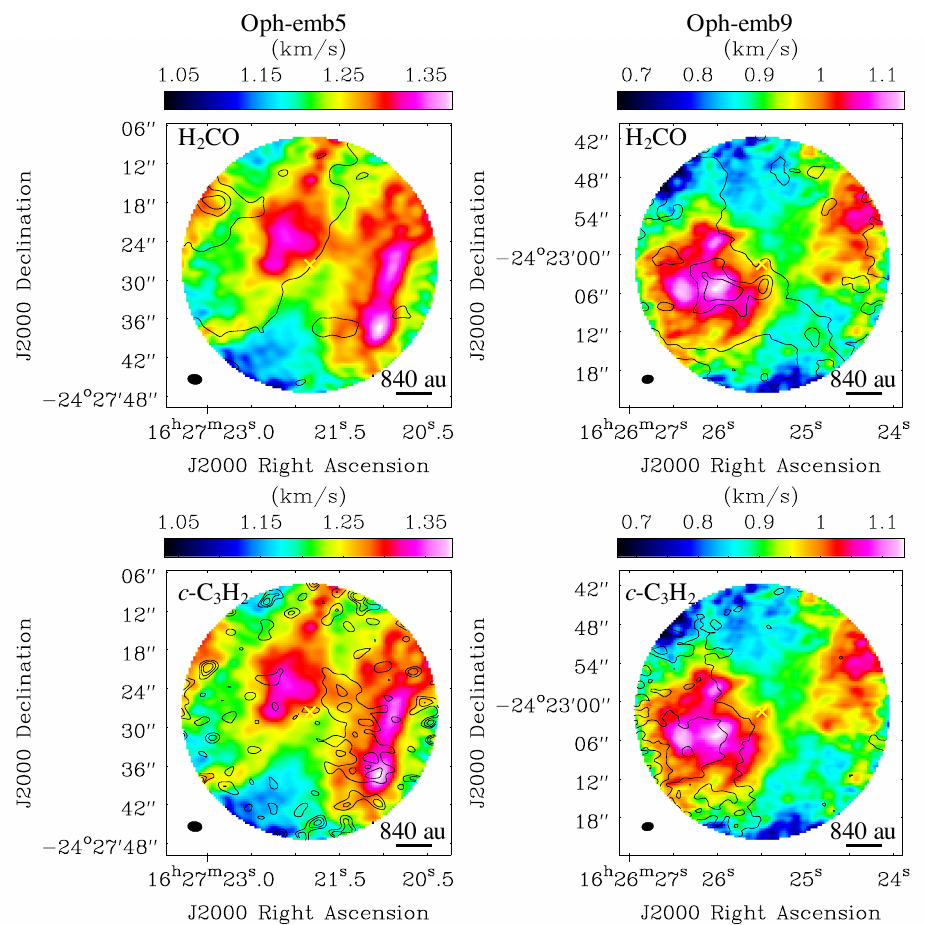}
 \end{center}
\caption{Moment 2 maps of the $^{13}$CO ($J=2-1$) line (color scales) overlaid by black contours showing moment 0 maps of the H$_{2}$CO ($3_{0, 3}-2_{0, 2}$) line (upper panels) and $c$-C$_{3}$H$_{2}$ ($6_{0, 6}- 5_{1, 5}$) line (bottom panels), respectively. Left and right panels show results in Oph-emb5 and Oph-emb9, respectively. The contour levels of the moment 0 images are the same as Figures \ref{fig:EES2009mom0} and \ref{fig:GY92mom0}. The yellow crosses indicate positions of the infrared sources \citep{2009ApJ...692..973E}. The filled black ellipses indicate the angular resolutions of 2\farcs2 $\times$ 1\farcs5 and 1\farcs8 $\times$ 1\farcs3 for Oph-emb5 and Oph-emb9, respectively. \label{fig:mom2_mom0}}
\end{figure*}

In this subsection, we investigate the relationship between the chemical differentiation and velocity dispersion, which is an indicator of gas turbulent motions.
Figure \ref{fig:mom2_mom0} shows comparisons of the $^{13}$CO moment 2 map (color scales) and the spatial distributions of H$_{2}$CO and $c$-C$_{3}$H$_{2}$ (black contours).
The black contours in the upper and bottom panels indicate moment 0 maps of the H$_{2}$CO ($3_{0, 3}-2_{0, 2}$) line and the $c$-C$_{3}$H$_{2}$ ($6_{0, 6}- 5_{1, 5}$) line, respectively.

There is no spatial relationship between the $^{13}$CO moment 2 map and the H$_{2}$CO moment 0 map around Oph-emb5, while the $c$-C$_{3}$H$_{2}$ moment 0 map peaks at the location of the largest velocity dispersion position.
The H$_{2}$CO emission may be associated with another source, which is located out of the field of view of these observations.
However, we cannot determine from the current data set.

The peak of the H$_{2}$CO moment 0 map corresponds to the largest velocity dispersion position in the $^{13}$CO moment 2 map around Oph-emb9.
These results may suggest that COMs (H$_{2}$CO and CH$_{3}$OH) are enhanced by the shock chemistry induced by the nearby Herbig Be star or the molecular outflow originated from the Oph-emb9 YSO \citep{2015MNRAS.447.1996W}.
Large-scale enhancement of CH$_{3}$OH by shock has been reported in other cluster-forming regions \citep[e.g.,][]{2020MNRAS.493.2395T}.
The H$_{2}$CO emission is strongly enhanced in the eastern direction from the IR source, while the enhancement is not efficient on the opposite side.
If the molecular outflow is the sole source of the enhancement of these COMs, this spatial difference cannot be explained. 
An alternative explanation is that a strong shock region at the eastern position is produced by a combination of the molecular outflow from Oph-emb9 and the effect of gas compression from the Herbig Be star. %"make clear "impact"

The peak in the $c$-C$_{3}$H$_{2}$ moment 0 map does not coincide with the peak in the $^{13}$CO moment 2 map.
As mentioned in Section \ref{sec:d2}, the PDR chemistry seems to enhance $c$-C$_{3}$H$_{2}$ around Oph-emb9, and the shock chemistry is unlikely related to the formation of $c$-C$_{3}$H$_{2}$.
In fact, the line widths of H$_{2}$CO and CH$_{3}$OH are larger than that of $c$-C$_{3}$H$_{2}$ (see Table \ref{tab:CASSIS}).
In summary, the chemical processes around Oph-emb9 are strongly affected by the nearby Herbig Be star, the UV radiation and probably gas compression.

\section{Conclusions} \label{sec:con}

We present Band 6 ALMA Cycle 4 archival data toward two YSOs in the Ophiuchus region. Oph-emb9 is irradiated by a nearby Herbig Be star, while Oph-emb5 is located in a relatively quiescent region.
Using the data, we investigate effects from a nearby bright star on the chemistry around the target YSOs.
The high angular resolution data with ALMA enable us to investigate the chemical processes in more detail compared to previous single-dish data \citep{2017ApJ...835....3L}.

We have detected $c$-C$_{3}$H$_{2}$, H$_{2}$CO, CH$_{3}$OH, $^{13}$CO, C$^{18}$O, DCO$^{+}$, and DCN in Oph-emb9, and all of them except for DCN in Oph-emb5.
The small hydrocarbon $c$-C$_{3}$H$_{2}$ and COMs (H$_{2}$CO and CH$_{3}$OH) show different spatial distributions around both YSOs, which indicate the chemical differentiation.

The $N$($c$-C$_{3}$H$_{2}$)/$N$(CH$_{3}$OH) column density ratios around Oph-emb9 are higher than those around Oph-emb5. 
Furthermore, $c$-C$_{3}$H$_{2}$ is greatly enhanced at its peak position, which is irradiated by the Herbig Be star.
These results indicate the PDR chemistry around the Oph-emb9 YSO driven by the Herbig Be star.

We compare the moment 0 maps of $c$-C$_{3}$H$_{2}$ and H$_{2}$CO with the $^{13}$CO moment 2 maps.
In the case of Oph-emb5, the peak position of the $c$-C$_{3}$H$_{2}$ moment 0 map corresponds to the position with the largest velocity dispersion in the $^{13}$CO moment 2 map.
Moreover, the peak of the H$_{2}$CO moment 0 map is consistent with the position with the largest velocity dispersion in Oph-emb9.
Such results imply that shock chemistry enhances the gas-phase COMs around Oph-emb9.
The strong shock seems to be induced by a combination of the nearby Herbig Be star and the molecular outflow from Oph-emb9 itself.

We have investigated the higher DCO$^{+}$ column density and the detection of DCN in Oph-emb9.
The main formation pathway of DCO$^{+}$ around Oph-emb9 seems to be the reaction HCO$^{+}$ + D $\rightarrow$ DCO$^{+}$ + H in relatively warm region ($30<T<100$ K).
This is supported by the detection of DCN, which is expected to be formed in warm regions.
In addition, the UV radiation from the Herbig Be star can enhance the gas-phase HCO$^{+}$ abundance with ion-molecule chemistry.
On the other hand, the dominant formation pathway of DCO$^{+}$ around Oph-emb5 is the reaction of H$_{2}$D$^{+}$ + CO $\rightarrow$ DCO$^{+}$ + H$_{2}$, which can proceed only in cold regions ($< 30$ K).

\begin{acknowledgments}
This paper makes use of the following ALMA data: ADS/JAO.ALMA\#2016.1.00319.S.
ALMA is a partnership of ESO (representing its member states), NSF (USA) and NINS (Japan), together with NRC (Canada), MOST and ASIAA (Taiwan), and KASI (Republic of Korea), in cooperation with the Republic of Chile. 
The Joint ALMA Observatory is operated by ESO, AUI/NRAO and NAOJ. 
This work included analyses carried out with the CASSIS software and the CDMS and JPL spectroscopic databases. 
CASSIS has been developed by IRAP-UPS/CNRS (http://cassis.irap.omp.eu).
This work was supported by JSPS KAKENHI grant No. JP20K14523. 
K.T. acknowledges support from Japan Foundation for Promotion of Astronomy.
This research was carried out in part at the Jet Propulsion Laboratory, which is operated for NASA by the California Institute of Technology. 
E.H. thanks the National Science Foundation for support through grant AST-1906489.
We thank the anonymous referee whose valuable comments helped improve the quality of the paper.
\end{acknowledgments}

\vspace{5mm}
\facilities{Atacama Large Millimeter/submillimeter Array (ALMA)}


\software{Common Astronomy Software Applications package \citep[CASA;][]{2007ASPC..376..127M}, CASSIS \citep{2011IAUS..280P.120C}} 



\appendix

\section{Channel maps of $^{13}$CO and C$^{18}$O lines} \label{sec:appendix1}

\begin{figure*}[!th]
\figurenum{11}
 \begin{center}
  \includegraphics[bb = 0 20 943 762, scale=0.5]{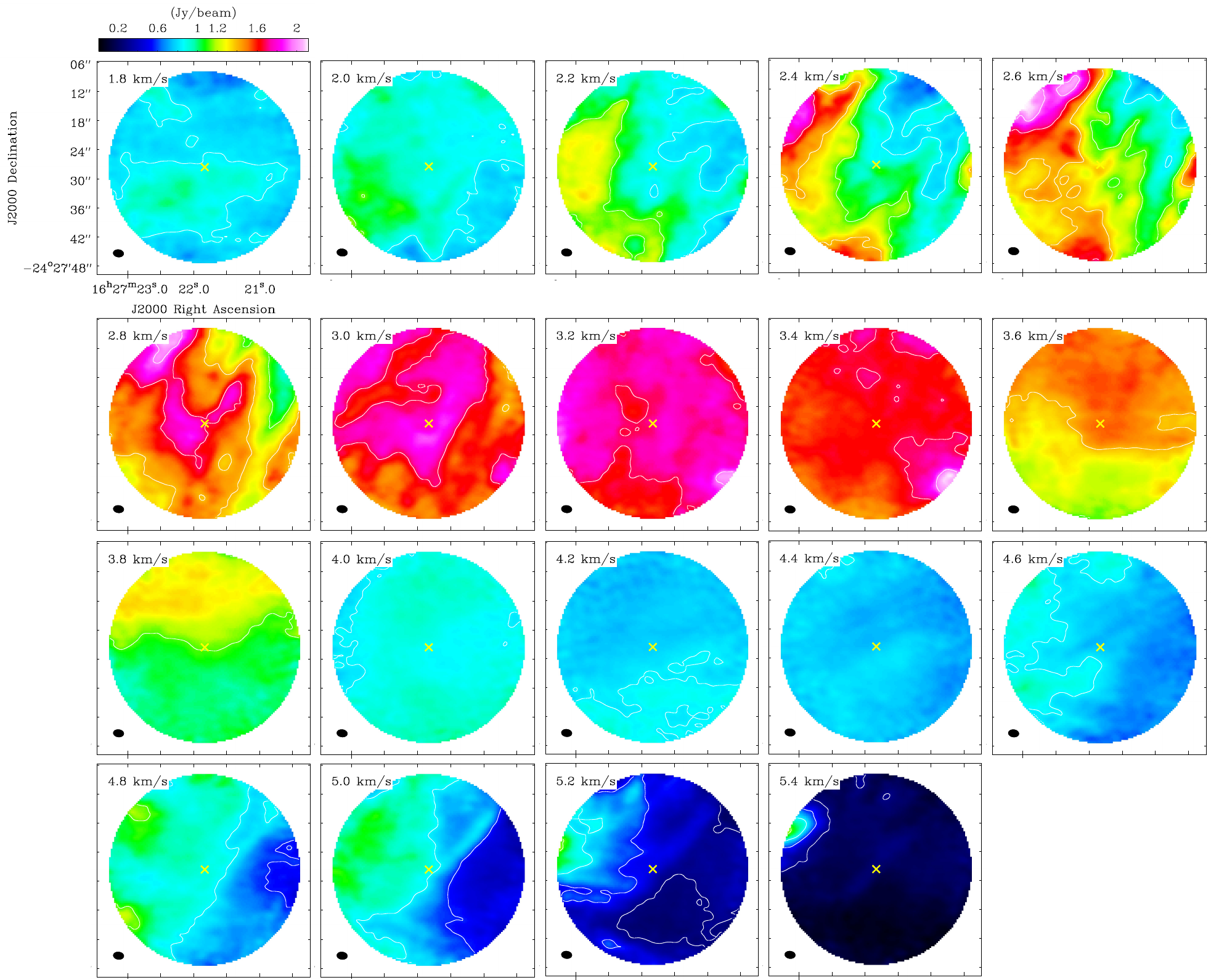}
 \end{center}
\caption{Channel maps of $^{13}$CO ($J=2-1$) toward Oph-emb5. Contour levels are in steps of $50\sigma$ from $50\sigma$ to $350\sigma$ ($1\sigma=4.5$ mJy\,beam$^{-1}$). The yellow crosses indicate positions of the infrared sources \citep{2009ApJ...692..973E}. The filled black ellipses indicate the beam sizes of 2\farcs2 $\times$ 1\farcs5. \label{fig:EES2009_13CO_channel}}
\end{figure*}

\begin{figure*}[!th]
\figurenum{12}
 \begin{center}
  \includegraphics[bb = 0 20 947 589, scale=0.5]{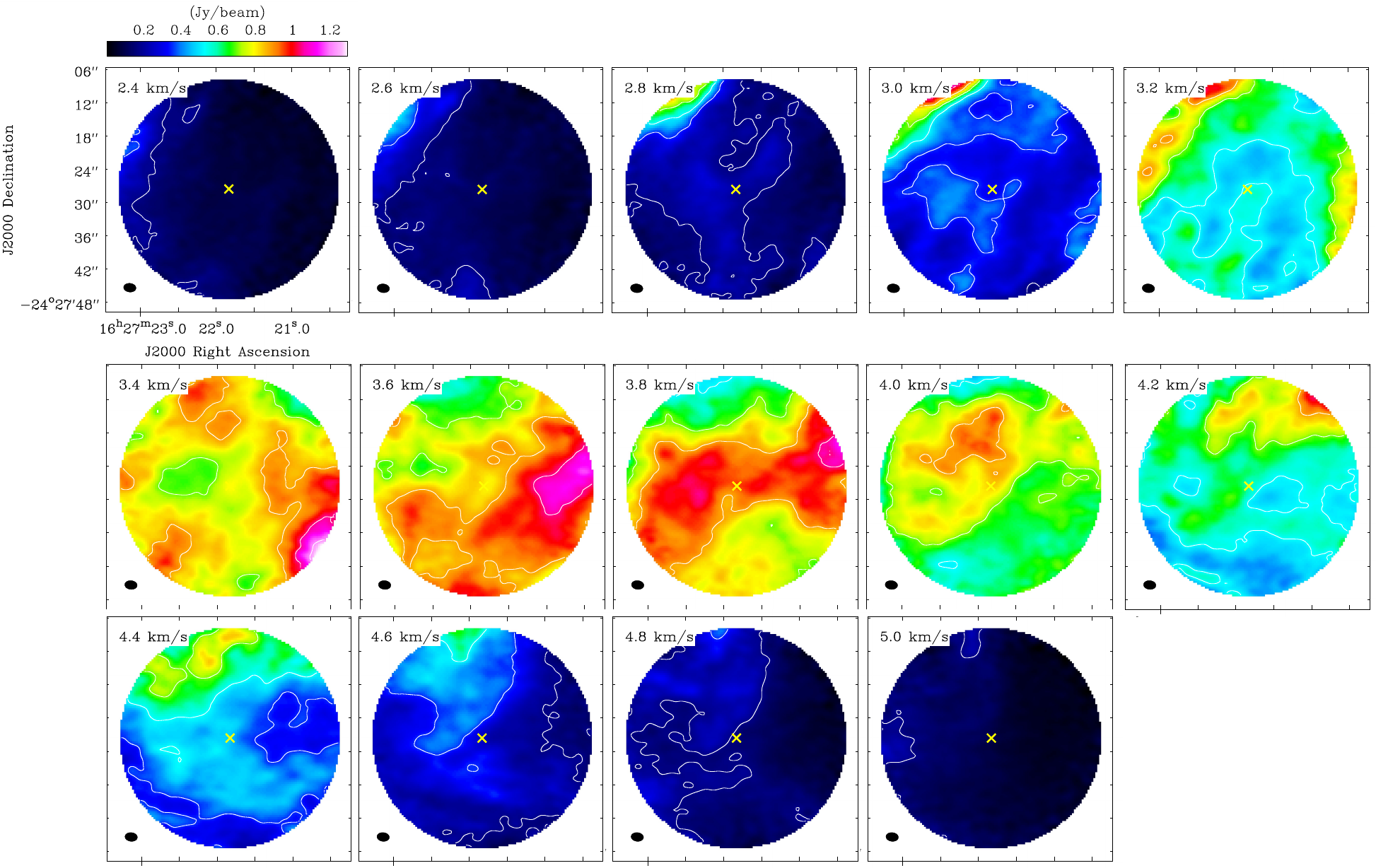}
 \end{center}
\caption{Channel maps of C$^{18}$O ($J=2-1$) toward Oph-emb5. Contour levels are in steps of $50\sigma$ from $50\sigma$ to $350\sigma$ ($1\sigma=3.0$ mJy\,beam$^{-1}$). The yellow crosses indicate positions of the infrared sources \citep{2009ApJ...692..973E}. The filled black ellipses indicate the beam sizes of 2\farcs2 $\times$ 1\farcs5. \label{fig:EES2009_C18O_channel}}
\end{figure*}

\begin{figure*}[!th]
\figurenum{13}
 \begin{center}
  \includegraphics[bb = 0 20 936 764, scale=0.5]{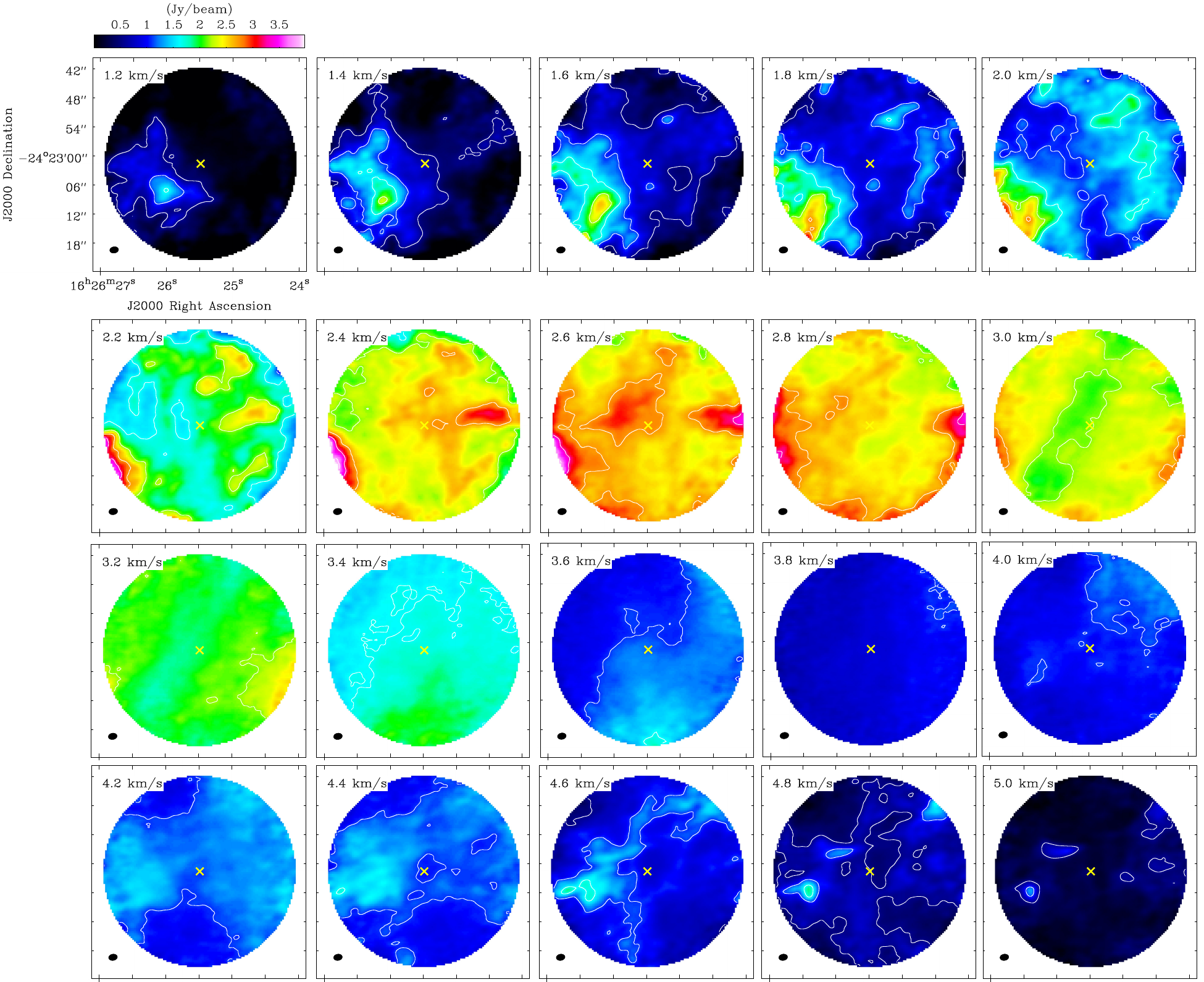}
 \end{center}
\caption{Channel maps of $^{13}$CO ($J=2-1$) toward Oph-emb9. Contour levels are in steps $50\sigma$ from $50\sigma$ to $350\sigma$ ($1\sigma=10$ mJy\,beam$^{-1}$). The yellow crosses indicate positions of the infrared sources \citep{2009ApJ...692..973E}. The filled black ellipses indicate the beam sizes of 1\farcs8 $\times$ 1\farcs3. \label{fig:GY92_13CO_channel}}
\end{figure*}

\begin{figure*}[!th]
\figurenum{14}
 \begin{center}
  \includegraphics[bb = 0 20 945 587, scale=0.5]{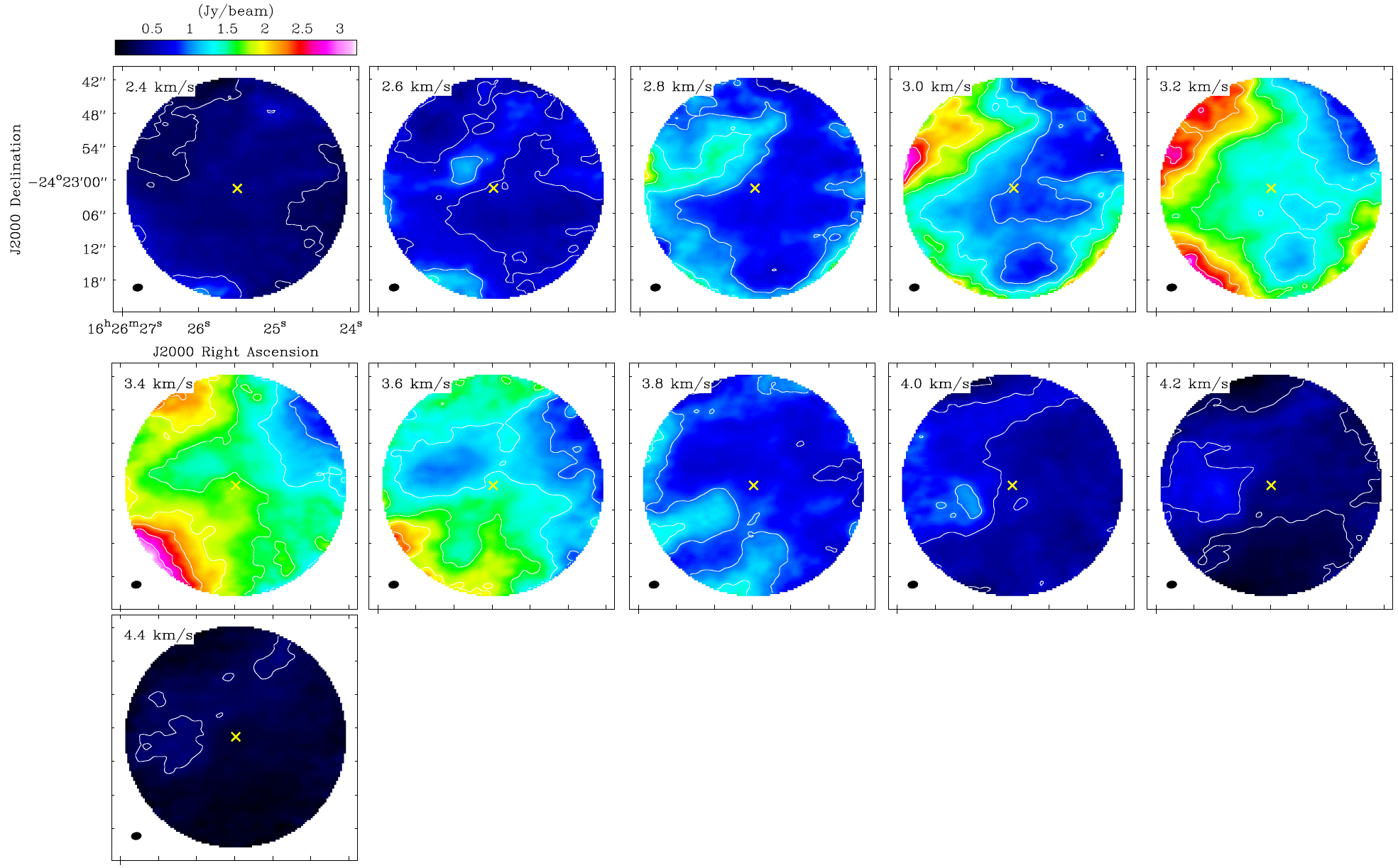}
 \end{center}
\caption{Channel maps of C$^{18}$O ($J=2-1$) toward Oph-emb9. Contour levels are in steps $50\sigma$ from $50\sigma$ to $450\sigma$ ($1\sigma=6.7$ mJy\,beam$^{-1}$). The yellow crosses indicate positions of the infrared sources \citep{2009ApJ...692..973E}. The filled black ellipses indicate the beam sizes of 1\farcs8 $\times$ 1\farcs3. \label{fig:GY92_C18O_channel}}
\end{figure*}

Figures \ref{fig:EES2009_13CO_channel}--\ref{fig:GY92_C18O_channel} show channel maps of the $^{13}$CO and C$^{18}$O lines towards Oph-emb5 and Oph-emb9, respectively.

\end{document}